\documentclass{elsart}
\usepackage{epsfig}
\usepackage{graphicx}

\newcommand{\app}{\mbox{${\rm \overline{p}}$ }}
\newcommand{\ps}{\mbox{${\rm e^{+}}$ }}

\begin{document}

\begin{frontmatter}



\title{PAMELA - A Payload for Antimatter Matter Exploration and Light-nuclei Astrophysics}


\author[a]{P.~Picozza},
\author[b]{A.M.~Galper},
\author[c]{G.~Castellini},
\author[d]{O.~Adriani},
\author[a] {F.~Altamura},
\author[e]{M.~Ambriola},
\author[f]{G.C.~Barbarino},
\author[a]{A. Basili},
\author[g]{G.A.~Bazilevskaja},
\author[a]{R.~Bencardino},
\author[i]{M.~Boezio},
\author[h]{E.A.~Bogomolov},
\author[d]{L.~Bonechi},
\author[d]{M.~Bongi},
\author[j]{L.~Bongiorno},
\author[i]{V.~Bonvicini},
\author[e]{F.~Cafagna},
\author[f]{D.~Campana},
\author[k]{P.~Carlson},
\author[a]{M.~Casolino},
\author[e]{C.~De Marzo\corauthref{dec}}, \corauth[dec]{Deceased.}
\author[a]{M.P.~De~Pascale},
\author[f]{G.~De Rosa},
\author[d]{D.~Fedele},
\author[k]{P.~Hofverberg},
\author[b]{S.V.~Koldashov},
\author[h]{S.Yu.~Krutkov},
\author[g]{A.N.~Kvashnin},
\author[k]{J.~Lund},
\author[i]{J.~Lundquist},
\author[g]{O.~Maksumov},
\author[a] {V.~Malvezzi},
\author[a] {L.~Marcelli},
\author[l]{W.~Menn},
\author[b]{V.V.~Mikhailov},
\author[a] {M.~Minori},
\author[g]{S.~Misin},
\author[i]{E.~Mocchiutti},
\author[a]{A.~Morselli},
\author[h]{N.N.~Nikonov},
\author[k]{S.~Orsi},
\author[f]{G.~Osteria},
\author[d]{P.~Papini},
\author[k]{M.~Pearce\corauthref{cor2}}, \corauth[cor2]{Corresponding author. e-mail:
pearce@particle.kth.se}
\author[j]{M.~Ricci},
\author[d]{S.B.~Ricciarini},
\author[b]{M.F.~Runtso},
\author[f]{S.~Russo},
\author[l]{M.~Simon},
\author[a]{R. Sparvoli\corauthref{cor}}, \corauth[cor]{Corresponding author. e-mail:
roberta.sparvoli@roma2.infn.it}
\author[d]{P.~Spillantini},
\author[g]{Yu.I.~Stozhkov},
\author[d]{E.~Taddei},
\author[i]{A.~Vacchi},
\author[d]{E.~Vannuccini},
\author[b]{S.A.~Voronov},
\author[b]{Y.T.~Yurkin},
\author[i]{G.~Zampa},
\author[i]{N.~Zampa},
\author[b]{V.G.~Zverev}
\address[a]{INFN, Structure of Rome ``Tor Vergata" and Physics
Department of University of Rome ``Tor Vergata", Via della Ricerca
Scientifica 1, I-00133 Rome, Italy}
\address[b]{Moscow Engineering and Physics Institute,
Kashirskoe Shosse 31, RU-115409 Moscow, Russia}
\address[d]{INFN, Structure of
Florence and Physics Department of University of Florence, Via
Sansone 1, I-50019 Sesto Fiorentino, Florence, Italy}
\address[c]{IFAC, Via Madonna del Piano 10, I-50019 Sesto Fiorentino, Florence, Italy}
\address[i]{INFN, Structure of Trieste and Physics
Department of University of Trieste, Via A. Valerio 2, I-34127
Trieste, Italy}
\address[k]{KTH, Department of Physics,
Albanova University Centre, SE-10691 Stockholm, Sweden}
\address[f]{INFN, Structure of Naples and Physics Department of University of Naples ``Federico II",  Via Cintia, I-80126 Naples, Italy}
\address[l]{Universit\"{a}t Siegen,  D-57068 Siegen, Germany}
\address[j]{INFN, Laboratori Nazionali di Frascati, Via Enrico Fermi 40,  I-00044 Frascati, Italy}
\address[e]{INFN, Structure of Bari and Physics
Department of University of Bari, Via Amendola 173, I-70126 Bari,
Italy}
\address[h]{Ioffe Physical Technical Institute,  Polytekhnicheskaya 26, RU-194021 St. Petersburg, Russia}
\address[g]{Lebedev Physical Institute, Leninsky Prospekt 53, RU-119991 Moscow, Russia}

\begin{abstract}

The PAMELA experiment is a satellite-borne apparatus designed to study charged
particles in the cosmic radiation with
 a particular focus on antiparticles. PAMELA is mounted on the Resurs DK1 satellite that was
 launched from the Baikonur cosmodrome on June 15$^{th}$ 2006. The PAMELA apparatus comprises
 a time-of-flight system, a magnetic spectrometer, a silicon-tungsten electromagnetic calorimeter,
 an anticoincidence system, a shower tail catcher scintillator and a neutron detector.
 The combination of these devices allows antiparticles to be reliably identified from a large
 background of other charged particles.
This paper reviews the design, space qualification and on-ground performance of PAMELA. The in-orbit performance will be discussed in future publications.

\end{abstract}

\end{frontmatter}

\section{Introduction}

The PAMELA (a Payload for Antimatter Matter Exploration and
Light-nuclei Astrophysics) experiment is a satellite-borne
apparatus designed to study charged particles in the cosmic
radiation with a particular focus on antiparticles (antiprotons
and positrons). PAMELA is installed inside a pressurized container
attached to a Russian Resurs DK1 earth-observation satellite that
was launched into space by a Soyuz-U rocket on June 15$^{th}$~2006 from the Baikonur cosmodrome in Kazakhstan. The
satellite orbit is elliptical and semi-polar, with an altitude
varying between 350~km and 600~km, at an inclination of
70$^{\circ}$. The mission is foreseen to last for at least three
years.

The PAMELA mission is devoted to the investigation of dark matter, the baryon asymmetry in the Universe, cosmic ray generation and propagation in our galaxy and the solar system, and studies of solar modulation and the interaction of cosmic rays with the earth's magnetosphere. The primary scientific goal is the study of the antimatter component of the cosmic radiation,
\begin{itemize}
\item in order to search for evidence of dark matter particle (e.g. non-hadronic particles outside the Standard Model) annihilations by precisely measuring the antiparticle (antiproton and positron) energy spectra;
\item in order to search for antinuclei (in particular, anti-helium);
\item in order to test cosmic-ray propagation models through precise measurements
of the antiparticle energy spectrum and precision studies of light nuclei
and their isotopes.
\end{itemize}
Concomitant goals include,
\begin{itemize}
\item a study of solar physics and solar
modulation during the 24$^{th}$ solar minimum by investigating low
energy particles in the cosmic radiation;
\item reconstructing the
cosmic ray electron energy spectrum up to several TeV thereby allowing a
possible contribution from local sources to be studied.
\end{itemize}
Table~\ref{t:science} shows
the design goals for PAMELA performance. The various cosmic-ray components and energy ranges over which
PAMELA will provide new results are presented.
\begin{table}   
\caption{\label{t:science} Design goals for PAMELA performance.}
\begin{center}
\begin{tabular}{|c|c|} \hline
{\bf Cosmic-ray particle}  & {\bf Energy range} \\ \hline Antiprotons &
80~MeV - 190~GeV \\ \hline Positrons & 50~MeV - 270~GeV \\ \hline
Electrons & 50~MeV - 400~GeV \\ \hline Protons & 80~MeV - 700~GeV \\
\hline Electrons+positrons & up to 2~TeV \\ \hline Light nuclei
(up to Z=6) & 100~MeV/n - 250~GeV/n \\ \hline \hline
{\bf Antinuclei} & {\bf Sensitivity 95\% C.L.} \\ \hline
Antihelium/helium ratio & of the
order of 10$^{-7}$  \\ \hline
\end{tabular}
\end{center}
\end{table}

\begin{figure}
\begin{center}
\epsfig{file=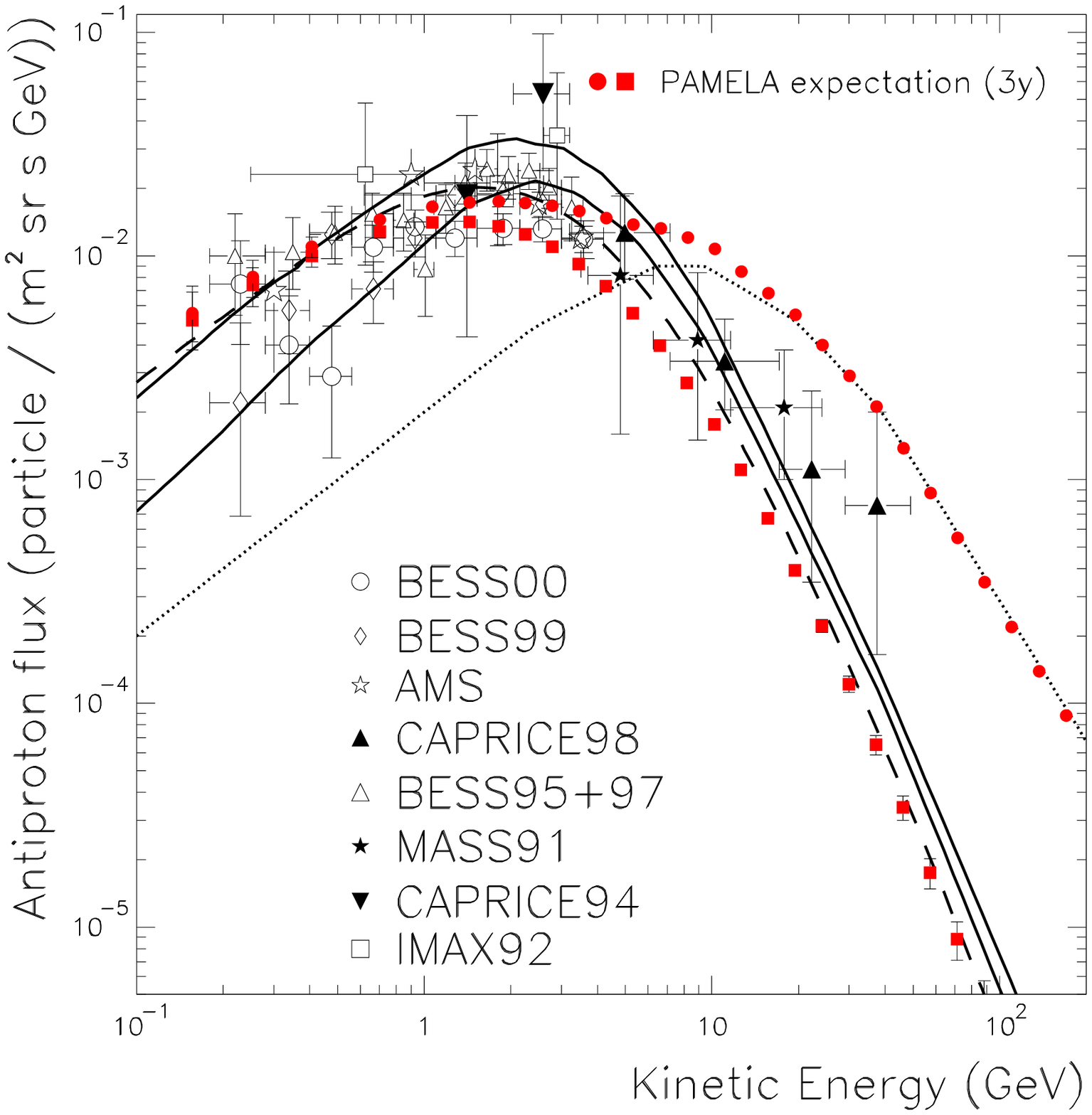, width=9cm} \caption{Recent
experimental \app spectra (BES\-S00 and BES\-S99~\cite{asa02},
AMS~\cite{agu02}, CAPRICE98~\cite{boe01a},
BES\-S95+97~\cite{ori00}, MASS91~\cite{bas99},
CAPRICE94~\cite{boe97}, IMAX92~\cite{mit96}) along with
theoretical calculations for pure \app secondary production (solid
lines: \cite{sim98}, dashed line: \cite{ber99b}) and for pure \app
primary production (dotted line: \cite{ull99}, assuming the annihilation of neutralinos of
mass 964~GeV/c$^2$). The expected
PAMELA performance, in case of a pure secondary component (full
boxes) and of an additional primary component (full circles), are
indicated. Only statistical errors are included in the expected
PAMELA data.} \label{pbflu}
\end{center}
\end{figure}

\begin{figure}
\begin{center}
\epsfig{file=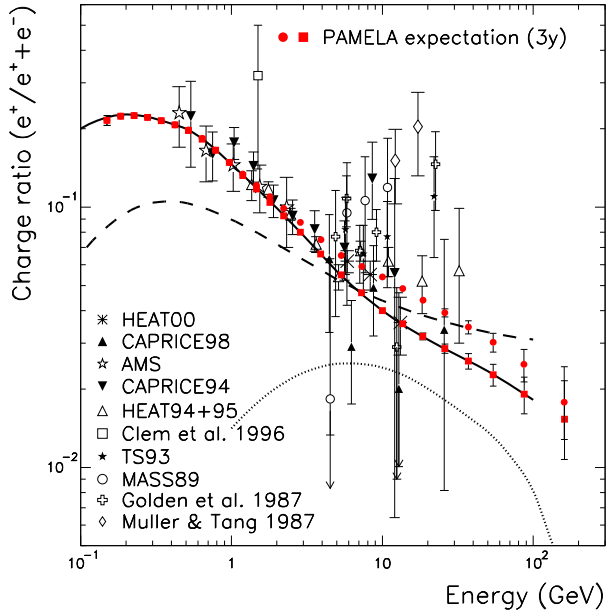, width=9cm} \caption{The positron
fraction as a function of energy measured by several experiments
(\cite{gol87,mul87,cle96} and MASS89~\cite{gol94},
TS93~\cite{gol96}, HEAT94+95~\cite{bar98}, CAPRICE94~\cite{boe00},
AMS~\cite{alc00}, CAPRICE98~\cite{boe01b}, HEAT00~\cite{bea04}).
The dashed \cite{pro82} and the solid \cite{mos98} lines are
calculations of the secondary positron fraction. The dotted line
is a possible contribution from annihilation of neutralinos of
mass 336~GeV/c$^2$ \cite{bal99}. The expected PAMELA performance, for a
pure secondary component (full boxes) and of an additional primary
component (full circles), are indicated. Only statistical errors are
included in the expected PAMELA data.} \label{posratio}
\end{center}
\end{figure}

Antiparticle measurements are the main scientific goal of the
experiment. The precise determination of the antiproton and
positron energy spectra will provide important information concerning
cosmic-ray propagation and solar modulation. For example,
indications of charge dependent solar modulation effects have been
already seen in the antiproton to proton ratio data~\cite{asa02}.
Antiparticles could also be produced from exotic sources such as
primordial black holes~\cite{mak96} or the annihilation of
supersymmetric~\cite{ber99a} or Kaluza-Klein~\cite{hoo05,bri05}
dark matter particles. Figures~\ref{pbflu} and~\ref{posratio} show
the current status of cosmic-ray antiproton and positron energy spectrum
measurements, respectively. Theoretical calculations for pure
secondary production~\cite{sim98,ber99b,pro82,mos98} and for pure
primary production due to the annihilation of supersymmetric dark
matter particles~\cite{ull99,bal99} are also shown. Almost all
data available so far have been obtained by balloon-borne
experiments.
PAMELA will be able to perform very precise
measurements with high statistics ($\sim$10$^4$ \app and
$\sim$10$^5$ \ps per year) and over a wider energy range than possible to date.
The full boxes in figures~\ref{pbflu} and
\ref{posratio} indicate the expected PAMELA performance in case of
a pure secondary antiproton and positron components and the full
circles show the expected performance in case of an additional
primary component. The errors on the expected PAMELA data points
only include statistical uncertainties. An average PAMELA orbit
has been used to estimate the vertical geomagnetic cut-offs and,
consequently, the expected number of antiproton and positron
events at low energies~\cite{lun04}.


\begin{figure}
\begin{center}
\epsfig{file=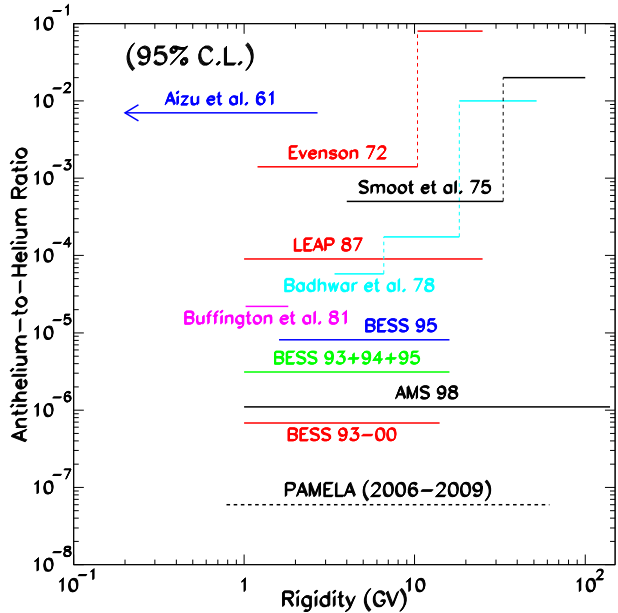, width=9cm}
\caption{The ratio of
anti-helium to helium in the cosmic radiation shown as a function
of rigidity~\cite{hebar}. No observation of antihelium has been made to date
and so upper limits are shown. The expectation for PAMELA after a
3~year long mission is shown.}
\label{antihe}
\end{center}
\end{figure}

Another prominent goal of PAMELA is to measure the
antihelium/helium ratio with a sensitivity of the order of
10$^{-7}$. This would represent a factor of 50~improvement on contemporary limits, as shown in figure~\ref{antihe} as a function of rigidity (momentum /charge)~\cite{hebar},\cite{antihe_flux}.
The contribution to the antihelium flux from cosmic ray interactions is expected to be less than
10$^{-12}$~\cite{antihe_flux} and so an observation of antihelium
would be a significant discovery as it could indicate the presence of antimatter domains in a baryon symmetric Universe.

The quasi-polar orbit and low geomagnetic cut-off experienced by
the PAMELA apparatus combined with its intrinsic ability to
measure low momenta will allow phenomena connected with solar and
earth physics to be investigated~\cite{cas04}.

The ability to measure the combined electron and positron
energy spectrum up to 2~TeV will allow the
contribution of local sources to the cosmic radiation to be
investigated (e.g. see~\cite{ato95}).

This article is organized as follows. The subdetector components
of the PAMELA instrument are discussed in section~\ref{sec:pamela}
along with results from performance studies with particle beams
and cosmic rays. The data acquisition and trigger systems are
described in section~\ref{sec:daq}. The Resurs DK1 satellite which
houses PAMELA is presented in section~\ref{sec:dk1}. Tests on
various qualification models of the PAMELA apparatus are
summarised in section~\ref{sec:qual}. The physics performance of
the flight instrument is presented in section~\ref{sec:ground}.

\section{The PAMELA apparatus}
\label{sec:pamela}

The PAMELA apparatus is composed of the following subdetectors,
arranged as shown in figure~\ref{pam}:

\begin{figure}
\begin{center}
\epsfig{file=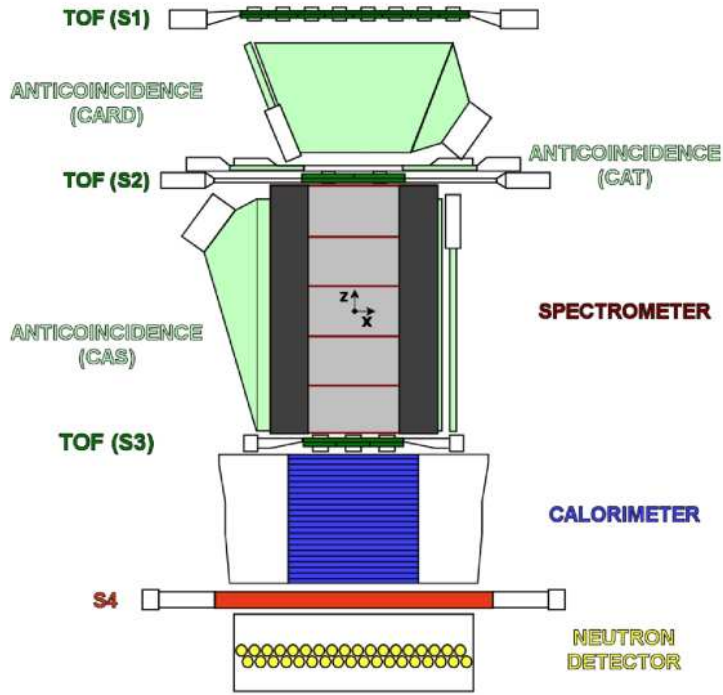,width=10cm}
\epsfig{file=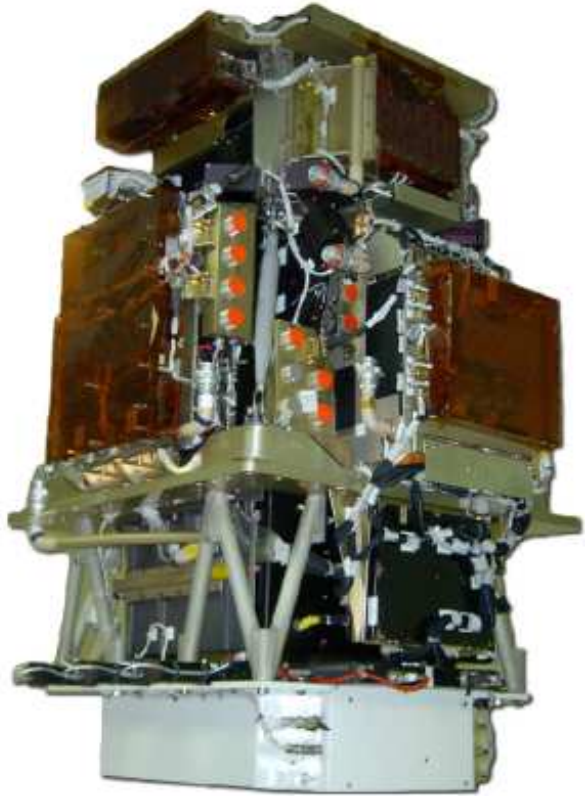, width=8cm}
\caption{The PAMELA instrument. Top: a schematic overview of the
apparatus. Bottom: a photograph taken just prior delivery to Russia
for integration with the Resurs DK1 satellite. The detector is
approximately 1.3~m tall. The magnetic field lines in the
spectrometer are oriented along the y~direction.} \label{pam}
\end{center}
\end{figure}

\begin{itemize}
    \item a time of flight system (ToF: S1, S2, S3);
    \item a magnetic spectrometer;
    \item an anticoincidence system (CARD, CAT, CAS);
    \item an electromagnetic calorimeter;
    \item a shower tail catcher scintillator (S4);
    \item a neutron detector.
\end{itemize}
The apparatus is $\sim$1.3~m high, has a mass of 470~kg and
an average power consumption of 355~W. The masses are
distributed according to table~\ref{tabmasse} and the power
consumption according to table~\ref{powertab}.

\begin{table}
\begin{center}
\caption{\label{tabmasse}  The PAMELA mass budget.}
\begin{tabular}{|l|c|}
\hline
{\bf Subsystem} & {\bf Mass (kg)}  \\
\hline
Spectrometer & 127 \\
Calorimeter         & 104 \\
General mechanics & 85\\
Electronic units & 45 \\
Neutron Detector & 30 \\
Thermal  system & 22 \\
Time-Of-Flight       & 18 \\
Anticoincidence     & 16 \\
Magnetic screens    & 15 \\
Bottom Scintillator        & 8 \\
 \hline
{\bf Total Mass} & {\bf 470} \\
 \hline
\end{tabular}
\end{center}
\end{table}
\begin{table}
\begin{center}
\caption{\label{powertab} The PAMELA average power budget.}
\begin{tabular}{|l|c|}
\hline
{\bf Subsystem} & {\bf Power (W)}  \\
\hline
Electronics & 80 \\
DC/DC converters & 74 \\
Spectrometer & 63 \\
Calorimeter         & 55 \\
CPU  & 35 \\
Power Supply system & 35 \\
Neutron Detector & 10 \\
Anticoincidence     & 1 \\
Bottom Scintillator        & 1 \\
Time-Of-Flight       & 1 \\
\hline
{\bf Total Power} & {\bf 355} \\
 \hline
\end{tabular}
\end{center}
\end{table}

\subsection{Overview}
\label{overview}

PAMELA is built around a 0.43~T permanent magnet spectrometer equipped with 6~planes of double-sided silicon
detectors allowing the sign, absolute value of charge and momentum of traversing charged particles to be determined.
 The acceptance of the spectrometer (which also defines the overall acceptance of the PAMELA experiment)
 is 21.5~cm$^2$sr and the maximum detectable rigidity is $\sim$1~TV.
 Spillover effects limit the upper detectable antiparticle momentum to $\sim$190~GeV/c ($\sim$270~GeV/c) for
 antiprotons (positrons). The spectrometer is surrounded by a plastic scintillator veto shield. An electromagnetic
 calorimeter mounted below the spectrometer measures the energy of incident electrons and allows topological
 discrimination between electromagnetic and hadronic showers (or non-interacting particles).
 Planes of plastic scintillator mounted above and below the spectrometer form a time-of-flight system which also
 provides the primary experimental trigger. The timing resolution of the time-of-flight system allows albedo
 particles to be identified and proton-electron separation is also possible below $\sim$1~GeV/c.
 Ionising energy loss measurements in the time-of-flight scintillator planes and the silicon planes of
  the magnetic spectrometer allow the absolute charge of traversing particles to be determined.
  The volume between the upper two time-of-flight planes is bounded by an additional plastic scintillator
  anticoincidence system. A plastic scintillator system mounted beneath the calorimeter aids in the identification of high energy electrons and is followed by a neutron detection system for the selection
  of very high energy electrons (up to 2~TeV) which shower in the calorimeter but do not necessarily pass through
  the spectrometer.

The PAMELA subdetectors are read out and controlled by a data
acquisition system based around Actel (54SX series) Field Programmable
Gate Arrays (FPGA)~\cite{pla} and Analog Devices (ADSP-2187L)
Digital Signal Processors (DSP)~\cite{dsp}. Connections between
different systems are realised with redundant data-strobe~\cite{DS} Low Voltage
Differential Signaling (LVDS) links. Each subdetector is also
connected to a global trigger system and can issue alarm
conditions (e.g. over-temperature, data corruption) to a
housekeeping system. All the data acquisition boards (except for
the calorimeter) are housed in a custom crate secured to the
PAMELA superstructure, as shown in figure~\ref{pam} (bottom). In order to
promote reliability, common design rules have been followed for
all electronics systems in PAMELA, e.g. over-current protection on
all electronics boards, redundant data links, redundant power
connections and the use of radiation qualified components.

\subsection{The Time of Flight (ToF) system}
\label{tof}

\begin{figure}
\begin{center}
\epsfig{file=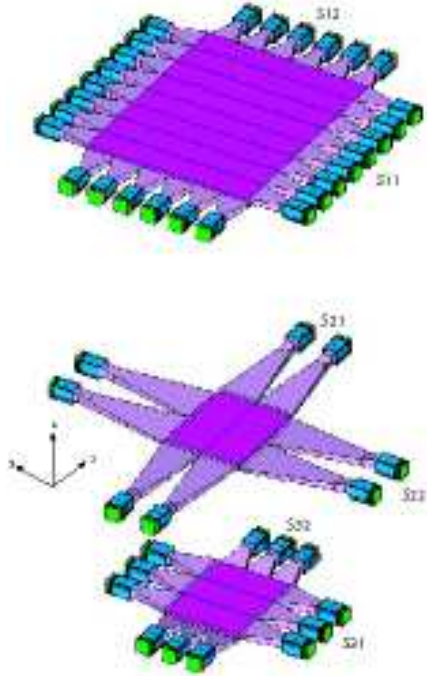, width=7cm} \caption{A schematic overview of the Time of Flight
system. The distance between the S1 and S3 planes is 77.3~cm.} \label{fig:tof}
\end{center}
\end{figure}

The ToF system~\cite{ost04a} comprises 6~layers of fast plastic scintillators
 (Bicron BC-404~\cite{bicron}) arranged in three planes (S1, S2 and S3),
 with alternate layers placed orthogonal to each other, as shown in figure~\ref{fig:tof}.
 The distance between S1 and S3 is 77.3~cm. Time-of-flight information for charged particles
 passing between planes S1 and S3 is combined with track length information derived from the magnetic
 spectrometer (see section~\ref{sec:spectrometer}) to determine particle velocities and reject albedo
 particles. Ionisation (dE/dx) measurements in the scintillator layers allow the particle charge to be
 determined at least up to Z$=$8. Coincidental energy deposits in combinations of planes provide the main trigger
 for the experiment, as described in section~\ref{sec:trigger}. The segmentation of each plane allows
 redundant studies of the trigger efficiency.

The sensitive area of each of the two S1 layers is
(33$\times$40.8)~cm$^{2}$ with the first layer divided into 8~bars
and the second layer divided into 6~bars. The total sensitive area
of the S2 and S3 planes is (15$\times$18)~cm$^{2}$ segmented
into 2$\times$2 and 3$\times$3 orthogonal bars, respectively. The
S1 and S3 layers are 7~mm thick while the S2 layers are 5~mm
thick. There are 24 scintillator bars in total. Both ends of each
scintillator bar are glued to a plastic light guide which is
mechanically coupled to a Hamamatsu R5900U photomultiplier (PMT)
by means of silicone pads of thickness 3~mm (S1 and S2) and 6~mm (S3). The differences in
thickness reflect the different vibrational spectra expected during launch.
The scintillators
and light-guides are wrapped in 2 layers of 25~$\mu$m thick Mylar
foil. The S3 plane is mounted directly on the base plate of
PAMELA, while the other two planes are enclosed in light-proof
boxes suspended off the PAMELA structure. A high-voltage divider
circuit is mounted directly behind each PMT. The high-voltage and
discrimination threshold for each PMT is chosen to optimize the
performance of a given ToF bar.

The ToF electronics system converts the 48~PMT pulses into time-
and charge-based measurements. In the timing section, a capacitor
is linearly charged during a time interval defined by the passage
of a particle through the ToF system. In the charge section, a
capacitor is charged with the PMT pulse charge. In both cases,
during read out the capacitor is linearly discharged into a
time-to-digital converter. The ToF electronics system comprises a
nine board electronics system based around the PAMELA-standard
FPGAs and DSPs. A separate trigger board processes
signals~\cite{ost04b} from the 48~PMTs as well as trigger signals
from the calorimeter and bottom scintillator (see section~\ref{sec:trigger}). Rate counters,
dead-/live-time counters and the logic to generate calibration
pulse sequences for different subsystems are also implemented.
Control masks select trigger types (see section 3.2) and allow
noisy or dead PMT channels to be vetoed and the PMT hit pattern to
be recorded for each trigger.

\begin{figure}
\begin{center}
\epsfig{file=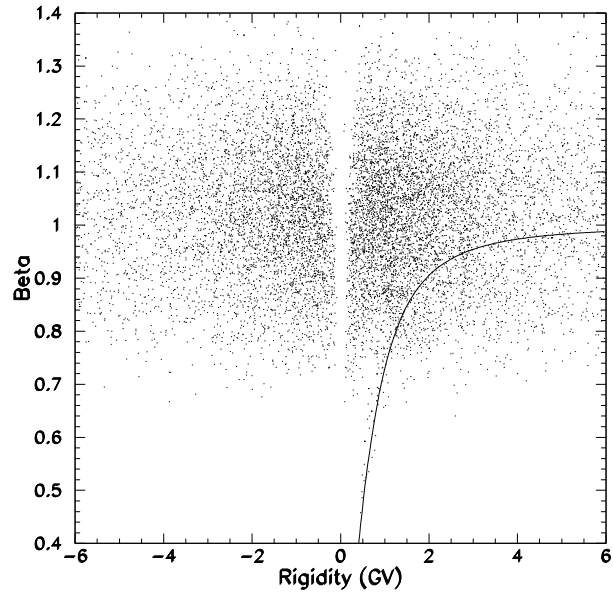, width=9cm}
\caption{The particle velocity measured by the ToF system as a function of rigidity.
The solid line is the theoretical $\beta$ for protons. The figure comprises 46000
events acquired with the final ToF system at ground.}
\label{tof1}
\end{center}
\end{figure}

\begin{figure}
\begin{center}
\epsfig{file=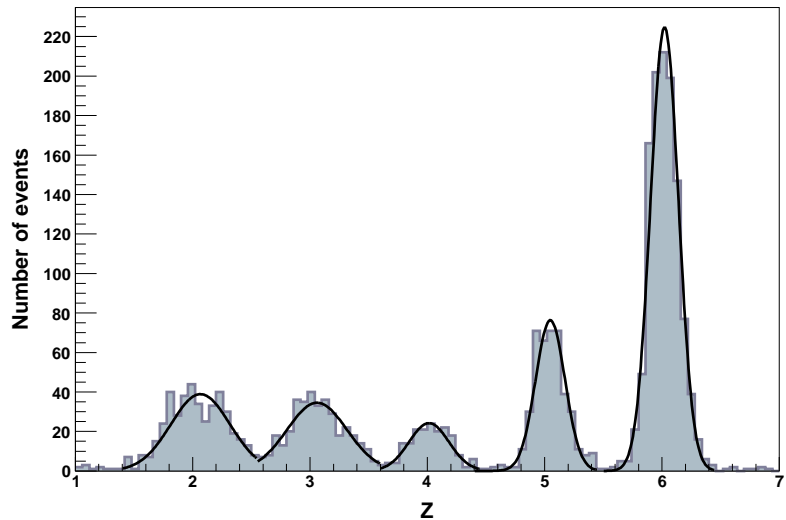,angle=270,width=9cm} \caption{Results from the GSI
beam test. Energy deposits in prototype S3 scintillators are converted by an
ADC to identify the secondaries produced by fragmentation
of the initial 1200 MeV/c $^{12}$C beam.}
\label{gsi}
\end{center}
\end{figure}

Figure~\ref{tof1} shows the velocity of particles (in units of speed of light, $\beta$)
measured by the ToF system as a function of their rigidity for data recorded at ground.
Most of the events are relativistic muons. A small proton component is visible at low
rigidity (the solid line indicates the theoretical $\beta$ for protons). The measured time-of-flight
 resolution of $\sim$250~ps will allow electrons (positrons) to be separated from antiprotons (protons) up
 to $\sim$1~GeV/c. Albedo particles can also be rejected with a significance of 60~standard deviations.
In addition, the measurement of ionization losses in the ToF
scintillators will allow the
 determination of the absolute charge of the particles, as shown in figure \ref{gsi}. These data were
 collected during a beam test performed at the GSI facility in Darmstadt.
Prototype versions of the S1, S2 and S3 ToF paddles were exposed
to $^{12}$C  beams. Targets of aluminium and polyethylene were used
to generate a variety of fragmentation products.
During this test, the S1 and S2 layers were
used to clean the data sample, and the particle charge was subsequently measured
using the S3 layer.
Data taken during this test also allowed the timing resolution for carbon to be determined as 70~ps.
This improvement is reasonable (compared to the 250~ps quoted above), since the timing resolution improves with the inverse square root of the number of photons created in the scintillator.

\subsection{Anticoincidence systems}
\label{ac}

\begin{figure}
\begin{center}
\mbox{\epsfig{file=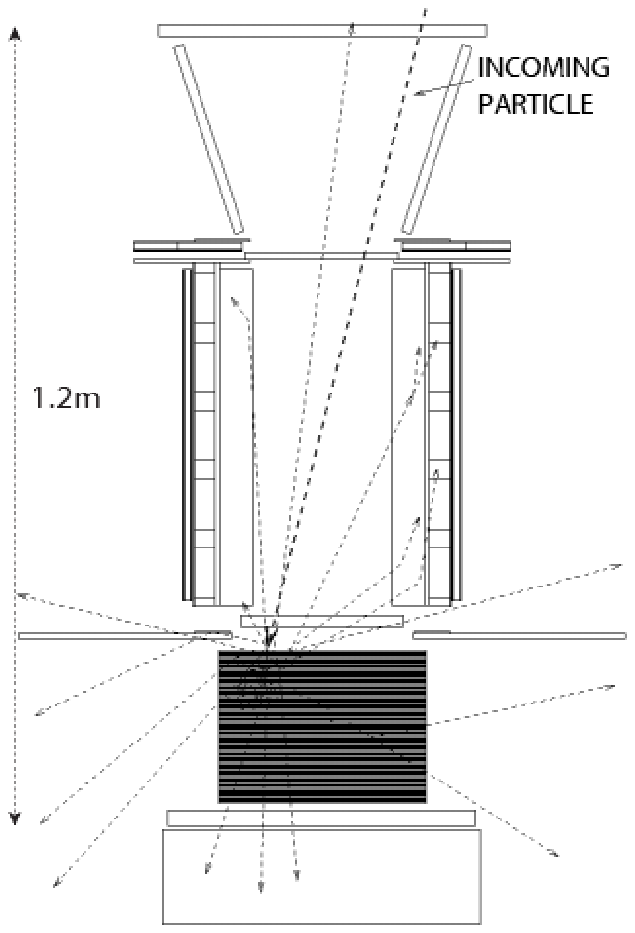, width=51mm} \hspace{-10mm} \epsfig{file=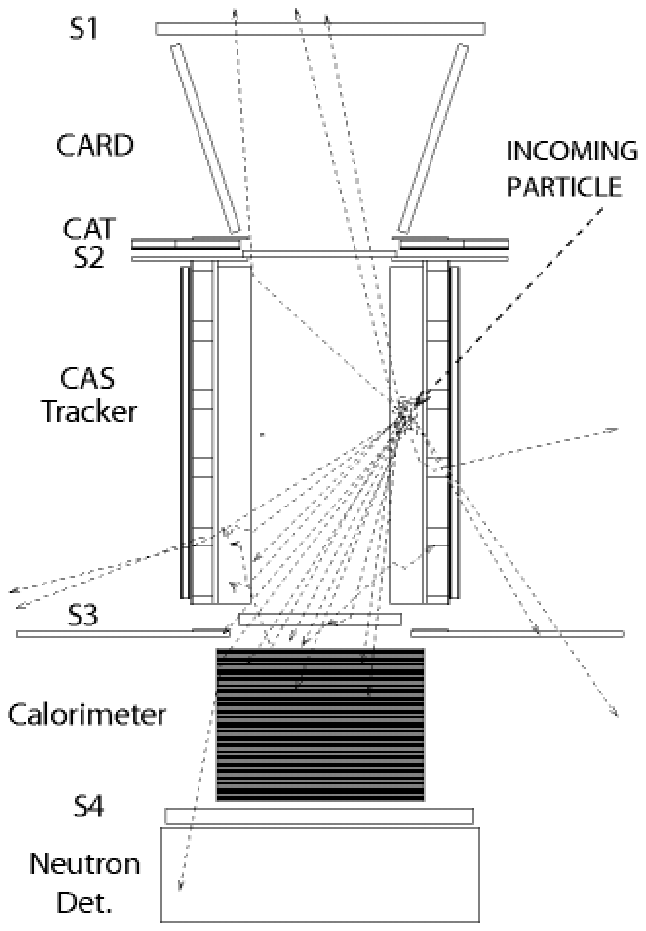, width=51mm}
 \hspace{-8mm}\epsfig{file=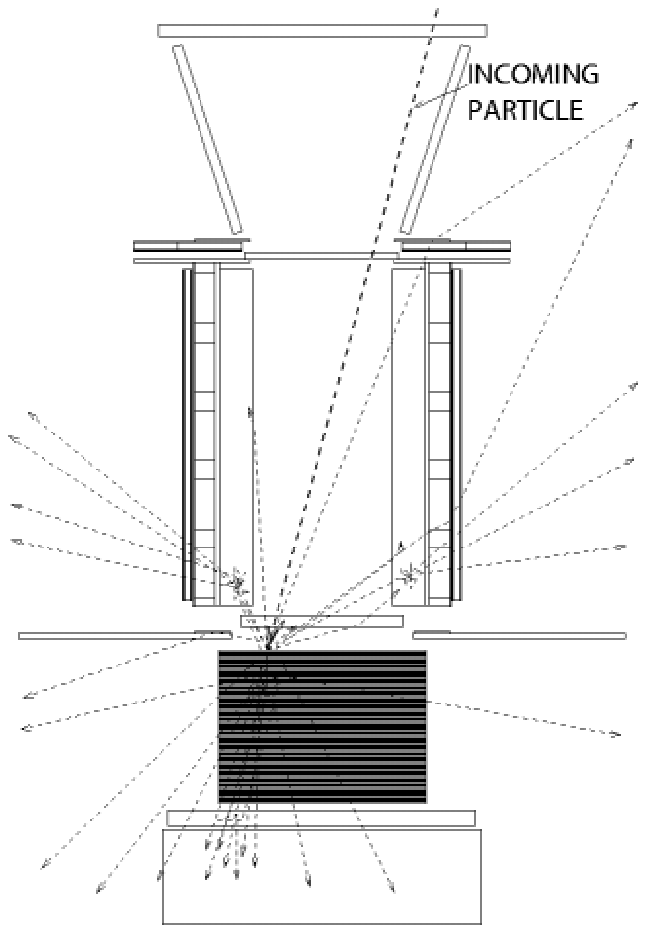, width=51mm}}
\caption{Schematic representations of simulated proton interactions in the PAMELA apparatus (non-bending view shown). Left: a good trigger
event without anticoincidence (AC) activity, with the lateral AC system (CAS) represented by the outermost rectangles bracketing the tracker. Centre: a false trigger created by a particle entering the apparatus from the side
 generating a shower and AC activity. Right: Particles backscattered from the calorimeter can also give rise to
 AC activity for good trigger events.}
\label{ints}
\end{center}
\end{figure}

Simulations have shown that the majority ($\sim$~75\%) of triggers in orbit are ``false" triggers~\cite{lun02}, i.e.
where the coincidental energy deposits in the time of flight scintillators are generated by secondary particles
produced in the mechanical structure of the experiment, as shown in figure~\ref{ints}. The aim of the
anticoincidence systems is to identify these events
during offline data analysis, or through the use of a second-level trigger in-orbit (see section~\ref{sec:trigger}).

\begin{figure}
\begin{center}
\epsfig{file=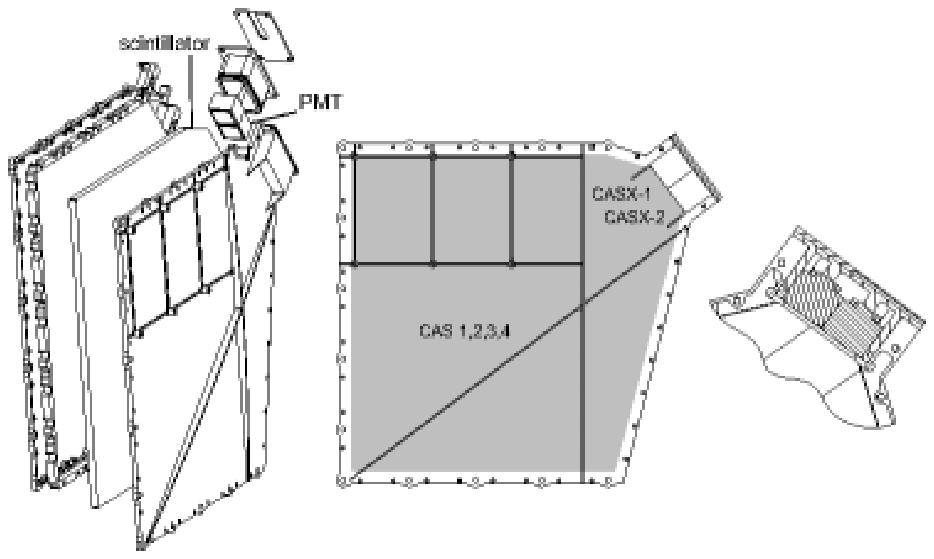, width=80mm}
\epsfig{file=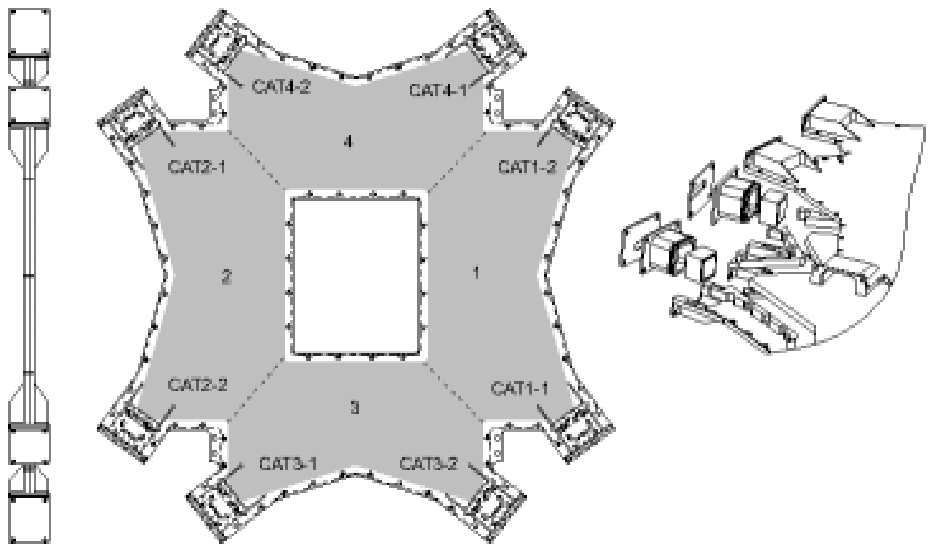, width=80mm}
\caption{An overview of the AC system. Top: the CAS system. Bottom: the CAT system. The CARD system is not
shown but the design closely follows that of CAS. The CAS scintillator is approximately 40~cm tall and 33~cm wide. The hole in the CAT scintillator measures approximately 22~cm by 18~cm.}
\label{fig:ac}
\end{center}
\end{figure}

The PAMELA experiment contains two anticoincidence (AC)
systems~\cite{ors05a}. The
primary AC system~\cite{ors04} consists of 4~plastic scintillators
(CAS) surrounding the sides of the magnet and one covering the top
(CAT), as shown in figure~\ref{fig:ac}. A secondary AC system consists of 4~plastic scintillators
(CARD) that surrounds the volume between the first two
time-of-flight planes. The CARD detectors are scaled-down versions of CAS. The AC systems use 8~mm thick plastic
scintillators (Bicron BC-448M~\cite{bicron}) read out by Hamamatsu~\cite{hama}
R5900U PMTs. Each scintillator is covered in two layers of
reflective Tyvek~\cite{tyvek} material and coupled via a 7~mm
thick silicone pad to the PMTs. Each CAS and CARD detector is read
out by two identical PMTs in order to decrease the possibility of
single point failure. Also for this reason, and to cover the
irregularly shaped area, the CAT detector is read out by 8 PMTs. A
high-voltage divider is mounted directly behind each PMT and
operated at a fixed voltage of -800~V. The scintillators and PMTs
are housed in aluminium containers which provide light-tightness,
allow fixation to the PAMELA superstructure and ensure that a
reliable scintillator-PMT coupling is maintained. The small fringe
field from the magnetic spectrometer at the position of the PMTs
means that additional magnetic shielding is not required.

The signals from the 24~PMTs are divided between two independent data acquisition boards with signals from PMTs for
 a given CAS or CARD detector or CAT quadrant routed to different boards. Only binary hit information is stored from each PMT indicating whether
 the deposited energy exceeds 0.5~mip (where 1~mip is the most probable energy deposited by a normally incident minimum ionising particle).
 On each board, an analogue front-end electronics system comprising an integration/amplification and discrimination
 stage processes the PMT signals before they are fed into a FPGA. The core of this digital system is a 16~bit shift
 register allowing hit information to be recorded in a time window of length 1.28~$\mu$s centered on the trigger time.
 Within this window the hit can be located with an accuracy of 80~ns.
The FPGA also allows the PMT singles rates to be monitored and
controls the data acquisition system. A DSP controls a
monitoring system which is based around 640~nm miniature LEDs
glued directly to the scintillator material.

The efficiency of the large area CAS detectors has been studied
using an external drift chamber to map the spatial distribution of
incident cosmic ray muons. A detection efficiency for mips of
(99.91$\pm$0.04)\% was observed~\cite{johan}. The AC system has
also been tested by studying the backscattering of particles (see
figure~\ref{ints}) from the calorimeter during tests with high
energy particle beams~\cite{level2}. The robustness of the AC
system has been determined by studying the stability of the
scintillator-PMT coupling to variations in
temperature~\cite{johan} and the vibration spectra expected
during launch~\cite{AC-vibe}.

\subsection{Magnetic spectrometer}
\label{sec:spectrometer}

The central part of the PAMELA apparatus is a magnetic
spectrometer~\cite{adr03} consisting of a permanent magnet and a
silicon tracker. The magnetic spectrometer is used to determine
the sign of charge and the rigidity of
particles up to $\sim$1~TV/c. Ionisation loss measurements are
also made in the silicon planes, allowing absolute particle charge
to be determined up to at least Z=6.

\begin{figure}
\begin{center}
\epsfig{file=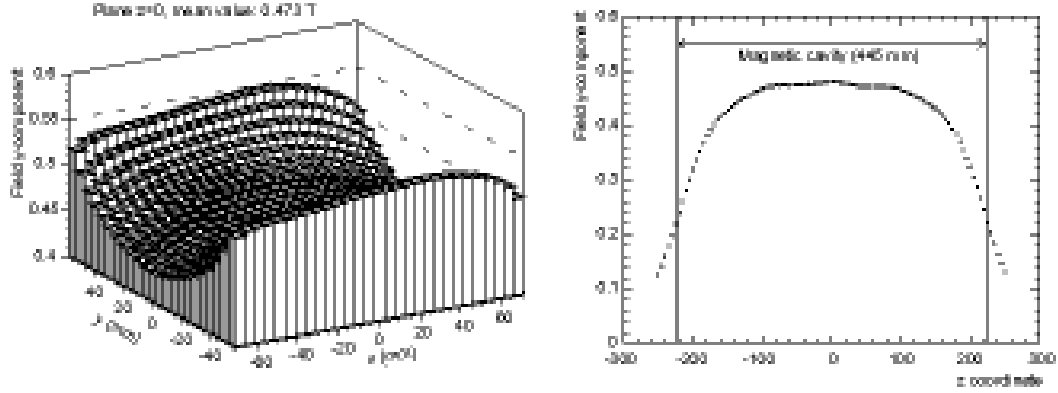, width=14cm} \caption{Left: the
y-component of the spectrometer magnetic field (T) measured at z=0.
Right: the variation of the y-component of the spectrometer
magnetic field (T) evaluated along the z-axis (mm).} \label{ybz0}
\end{center}
\end{figure}

The magnet is composed of five modules forming a tower 44.5~cm high. Each
module comprises twelve magnetic blocks, made of a Nd-Fe-B alloy with a
residual magnetisation of 1.3~T. The blocks are configured to provide an almost uniform
magnetic field oriented along the y-direction inside a cavity of dimensions
(13.1$\times$16.1)~cm$^2$. The dimensions of the permanent magnet define the
geometrical factor of the PAMELA experiment to be 21.5~cm$^2$sr. To allow precise rigidity
measurements to be obtained from the reconstructed particle trajectory, the
magnetic field has been precisely measured with a Hall probe through-out the cavity volume and the surrounding regions. Figure~\ref{ybz0} shows the y-component
of the magnetic field measured in the z=0 plane as a function of x
and y and the y-component as measured along the z-axis. The mean magnetic
field inside the cavity is 0.43~T with a value of 0.48~T measured at the centre.
Any stray magnetic field outside of the cavity can potentially interfere with
the satellite instruments and navigation systems. In order to attenuate the
stray field, the magnet is enclosed by ferromagnetic shielding.

\begin{figure}
\begin{center}
\epsfig{file=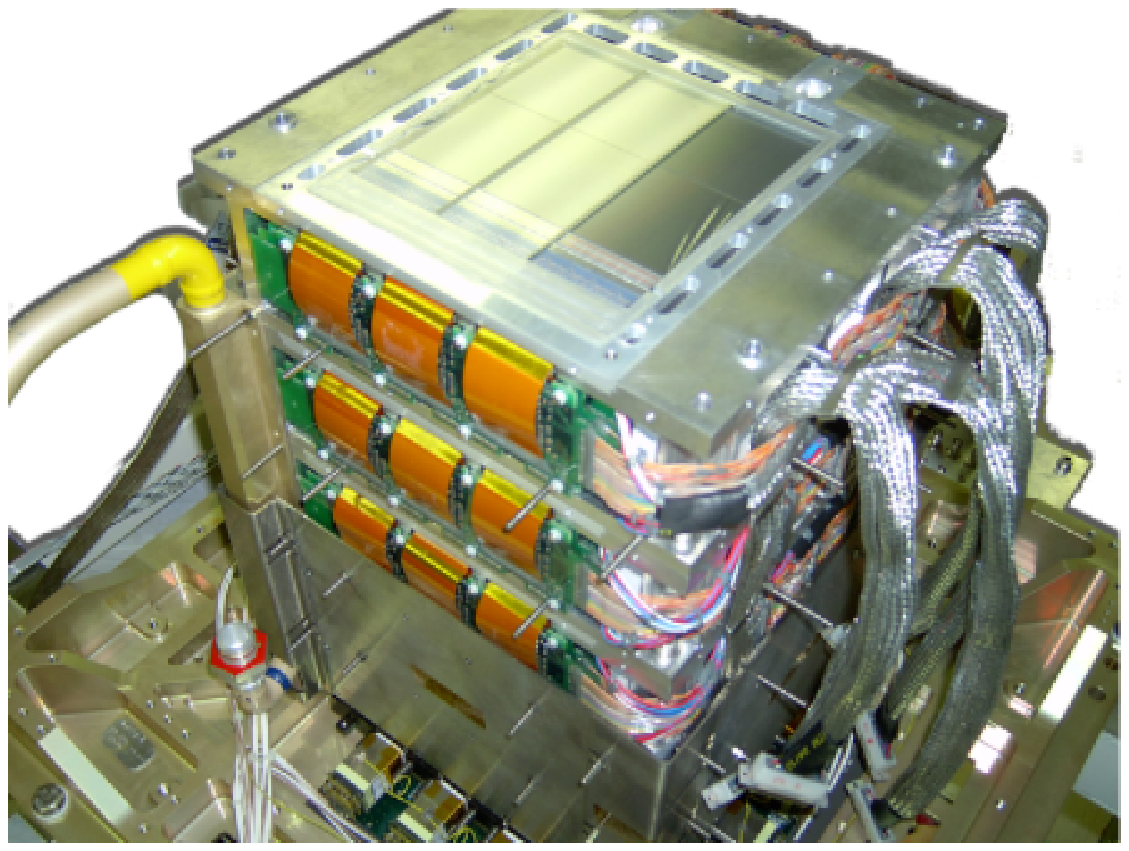,width=8cm}
\epsfig{file=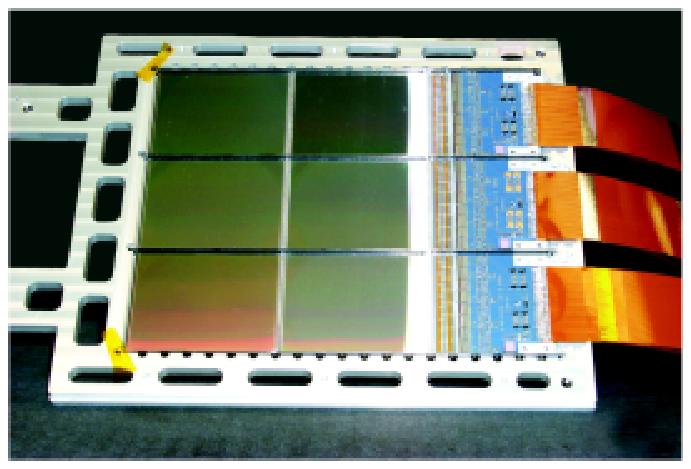, width=8cm}
\caption{Top: an overview of the magnetic spectrometer showing the top silicon plane.
The magnet cavity has dimensions (13.1$\times$16.1)~cm$^2$.
 A cooling loop enters from the left-hand side and the ADC boards mounted on the magnet canister are also visible. The lower part of the magnet canister is covered by a magnetic screen. Bottom: a
silicon plane comprising three silicon strip detectors and front-end
electronics.} \label{hybrid}
\end{center}
\end{figure}

Six equidistant 300~$\mu$m thick silicon detector planes are
inserted inside the magnetic cavity. The double-sided silicon
sensors provide two independent impact coordinates on each plane.
The basic detecting unit is the ladder which comprises two
sensors, (5.33$\times$7.00)~cm$^2$, assembled with a front-end
hybrid circuit, as shown in figure~\ref{hybrid}. Each plane is
built from three ladders that are inserted inside an aluminium
frame which connects to the magnet canister. In order to limit
multiple scattering in dead layers, no additional supporting
structure is present above or below the planes. Each high
resistivity n-type silicon detector is segmented into micro-strips
on both sides with p$^+$ strips implanted on the junction side
(bending-, x-view) and n$^+$ strips on the Ohmic side (non-bending, y-view). In
the x-view, the implantation pitch is 25~$\mu$m and the read-out
pitch is 50~$\mu$m. In the y-view, the read-out pitch is 67~$\mu$m
with the strips orthogonal to those in the x-view. The mip
efficiency for a single plane (including dead regions) exceeds 90\%.

The front-end electronics system is based around VA1 Application Specific Integrated Circuits (ASICs)~\cite{VA1} which contain 128~charge sensitive preamplifiers
connected to shapers and a sample and hold circuit. The
signals from the VA1 chips are sent over 5~cm long kapton cables
to be digitised by Analog-to-Digital (ADC) boards mounted on the magnet canisters. The digitised data are transferred by serial links to
DSP-based read-out boards where they are compressed using a Zero
Order Predictor (ZOP) algorithm. The compression factor is
estimated at 95\%.

The main task of the spectrometer is to measure the rigidity, $R$,
of charged particles. The momentum of the particle and the sign of
its electric charge can then be derived from the relation
$R=cp/Ze$, where $e$ is the electron charge, $p$ the momentum, $c$
the speed of light and $Z$ is the absolute charge. As discussed in section~\ref{tof}, the ToF system can be used to identify physical albedo activity. Instrumental albedo, e.g. due to particles backscattered from the calorimeter, can be identified a combination of the ToF system, the anticoincidence systems, and the tracking capabilities of the calorimeter (see section~\ref{calorimeter}).

\begin{figure}
\begin{center}
\epsfig{file=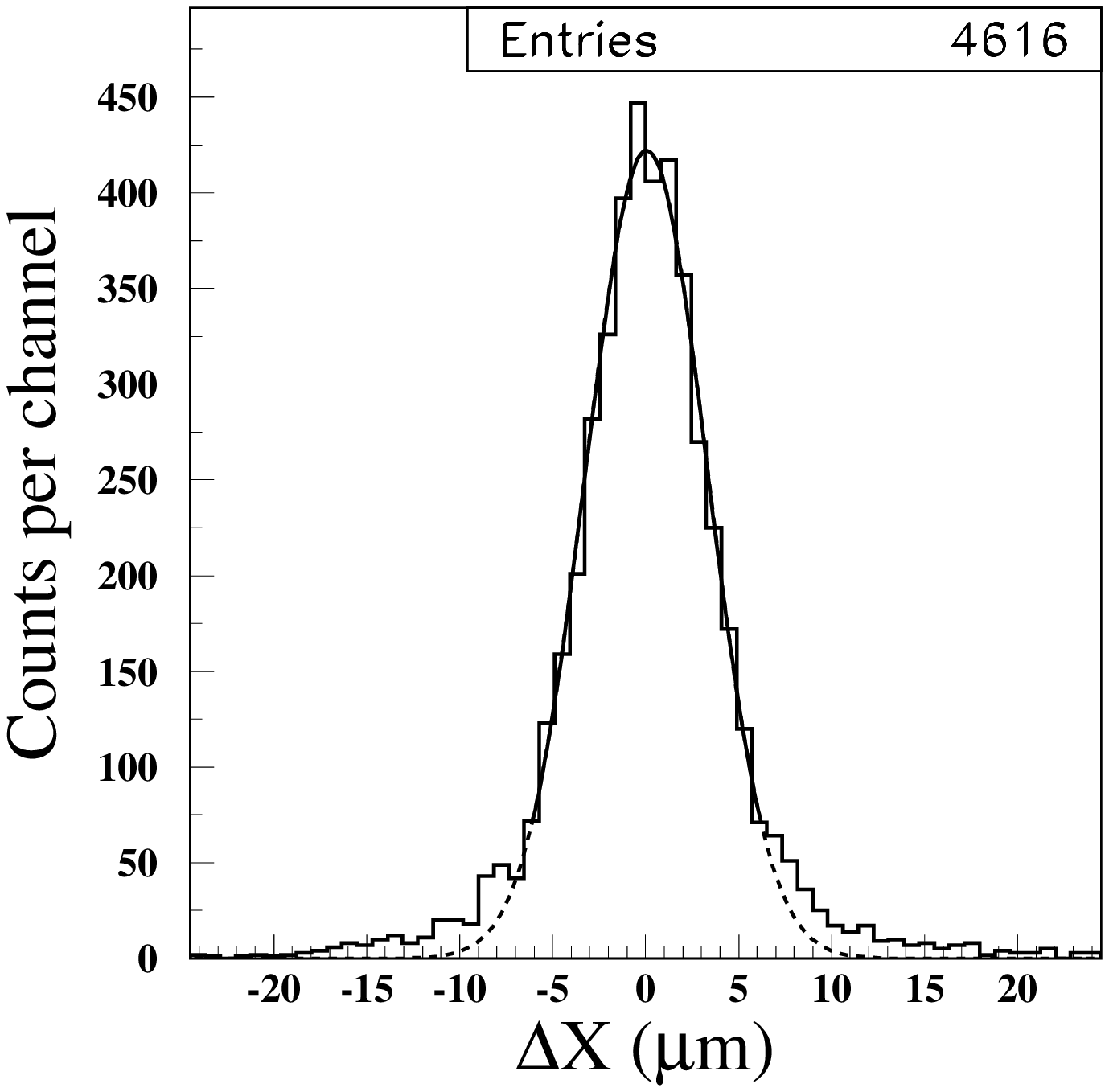,
width=6.5cm}\hspace{2mm}\epsfig{file=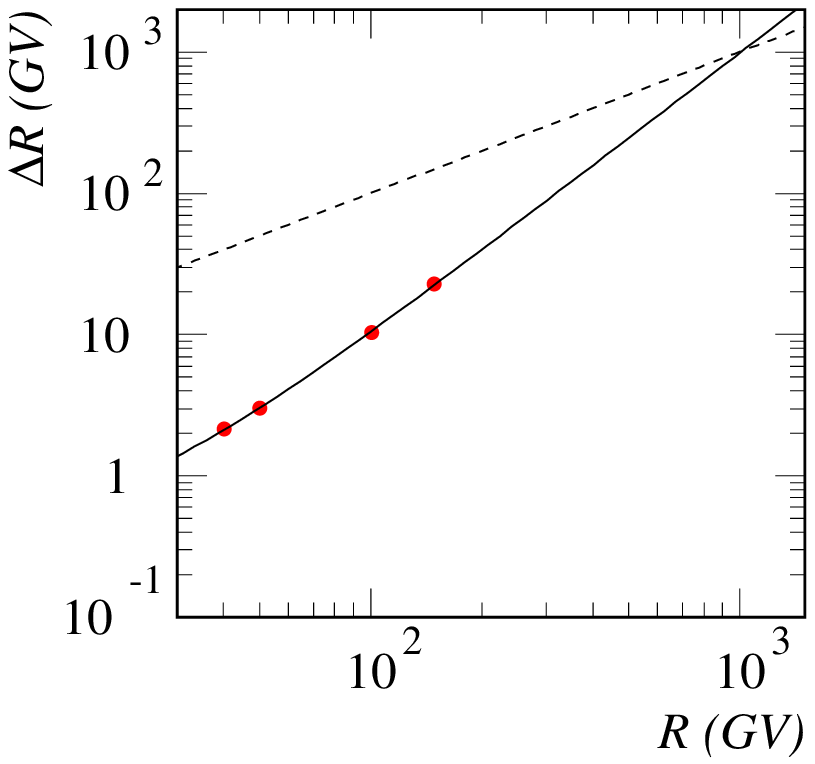,width=6.5cm}
\caption{Left: The spatial resolution of the tracker in the
bending view. The line indicates a Gaussian fit. Right: the
deflection error $\Delta R$ measured by the magnetic spectrometer
as a function of $R$ obtained with proton beams. The dashed line
is the bisector $\Delta R=R$. The functional form used to describe
the experimental $\Delta R$ curve is obtained by assuming that two
effects contribute to the deflection, $\eta$ = 1/R uncertainty,
namely multiple scattering and spatial resolution. The former can
be expressed (in the limit $\beta \sim$ 1) as $\Delta \eta_{ms}
\propto 1/R$. The latter is defined by $\Delta \eta_{res} = K$,
where $K$ is a constant. The intersection of the two curves gives
the maximum detectable rigidity of the spectrometer.} \label{res}
\end{center}
\end{figure}

The magnetic spectrometer measures the deflection of a particle, which is
defined as the inverse of the rigidity. The resolution in the deflection measurement
depends on the geometrical configuration of the spectrometer, on the intensity of the magnetic field and on the spatial resolution of the position measuring system - the silicon sensors in this case. This spatial resolution depends on the particle incidence angle. For normally
incident tracks, tests with particle beams show a spatial
resolution of (3.0$\pm$0.1)~$\mu$m and (11.5$\pm$0.6)~$\mu$m in the bending and non-bending
views, respectively. The spatial resolution in the bending view is shown in
figure~\ref{res} (left). Figure~\ref{res} (right) shows the resulting deflection error as a
function of rigidity obtained with proton beams. From this plot a maximum
detectable rigidity (MDR)\footnote{Defined as a 100\% uncertainty
in the rigidity determination.} of $\sim 1$~TV can be inferred. Note that this exceeds the design goal presented in table~\ref{t:science}. In
flight, the deflection measurement of the tracking system will be
cross-checked with the energy measurement of the calorimeter for
high-energy electrons.

\subsection{Electromagnetic Calorimeter}
\label{calorimeter}

Protons and electrons dominate the positively and negatively
charged components of the cosmic radiation, respectively. The main
task of the calorimeter is to select positrons and antiprotons
from like-charged backgrounds which are significantly more
abundant. Positrons must be identified from a background of
protons that increases from about 10$^3$ times the positron
component at 1~GeV/c to $\sim$5$\times$10$^3$ at 10~GeV/c,  and
antiprotons from a background of electrons that decreases from
$\sim$5$\times$10$^3$ times the antiproton component at 1~GeV/c to
less than 10$^2$ times above 10~GeV/c. This means that the PAMELA
system must separate electrons from hadrons at a level of 10$^5$ -
10$^6$. Much of this separation must be provided by the
calorimeter, i.e. electrons must be selected with an acceptable
efficiency and with as small a hadron contamination as possible.

\begin{figure}
\begin{center}
\epsfig{file=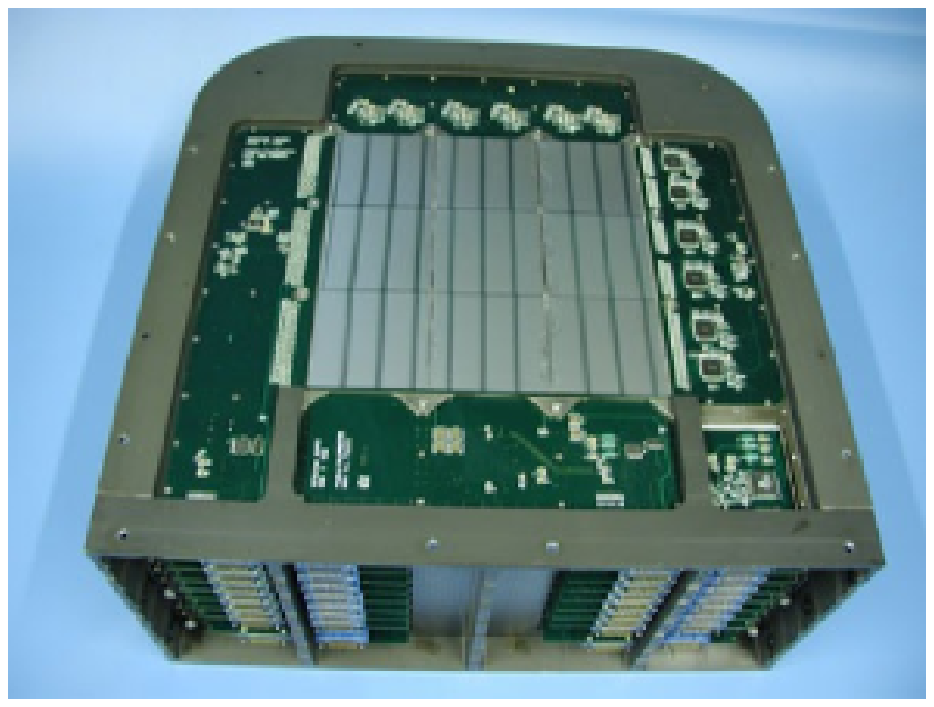, width=8cm}
\epsfig{file=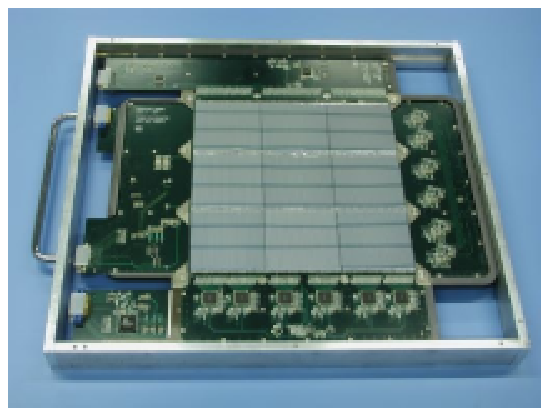, width=8cm}
\caption{Top: The PAMELA
electromagnetic calorimeter with the topmost silicon
plane visible. The device is $\sim$20~cm tall and the active
silicon layer is $\sim$24$\times$24~cm$^2$ in cross-section. Bottom: Detail of a single
calorimeter module comprising a tungsten layer sandwiched between two silicon detector planes.}
\label{calo}
\end{center}
\end{figure}

The sampling electromagnetic calorimeter comprises 44 single-sided
silicon sensor planes (380~$\mu$m thick) interleaved with
22~plates of tungsten absorber~\cite{boe02}. Each tungsten layer
has a thickness of 0.26~cm, which corresponds to 0.74~X$_0$
(radiation lengths), giving a total depth of 16.3~X$_0$
($\sim$0.6~nuclear interaction lengths). Each tungsten plate is
sandwiched between two printed circuit boards upon which the
silicon detectors, front-end electronics and ADCs are mounted. The
(8$\times$8)~cm$^2$ silicon detectors are segmented into
32~read-out strips with a pitch of 2.4~mm. The silicon detectors
are arranged in a 3$\times$3 matrix and each of the 32~strips is
bonded to the corresponding strip on the other two detectors in
the same row (or column), thereby forming 24~cm long read-out
strips. The orientation of the strips of two consecutive layers is
orthogonal and therefore provides two-dimensional spatial
information (``views"). Figure~\ref{calo} shows the calorimeter
prior to integration with the other PAMELA detectors.

The calorimeter front-end electronics is based around the CR1.4P
ASIC~\cite{cr14p} which provides 16~channels containing a
charge-sensitive preamplifier, a CR-RC shaper, a track-and-hold circuit and an
output multiplexer. A charge-injection calibration system is also
implemented. Six CR1.4P chips are used per plane with the outputs
multiplexed into a single 16-bit ADC. Data from all 44~ADCs are
processed by 4 DSP-based read-out boards mounted within the
calorimeter housing before being sent over serial links to the
main PAMELA data acquisition system. The read-out is divided into
4 independent sections, corresponding to the x-even, y-even, x-odd
and y-odd planes.

\begin{figure}
\begin{center}
\epsfig{file=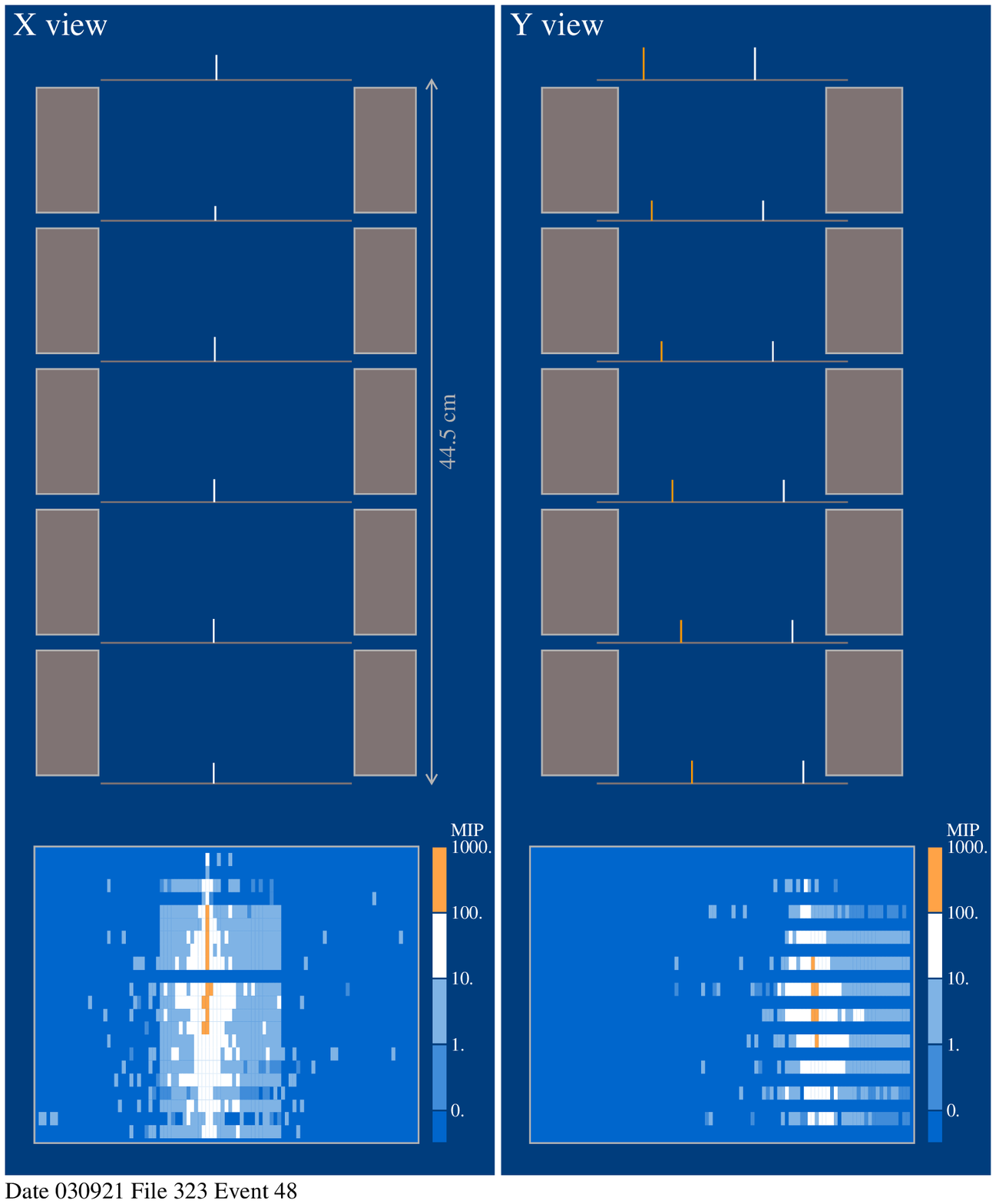,
width=6.5cm}\hspace{2mm}\epsfig{file=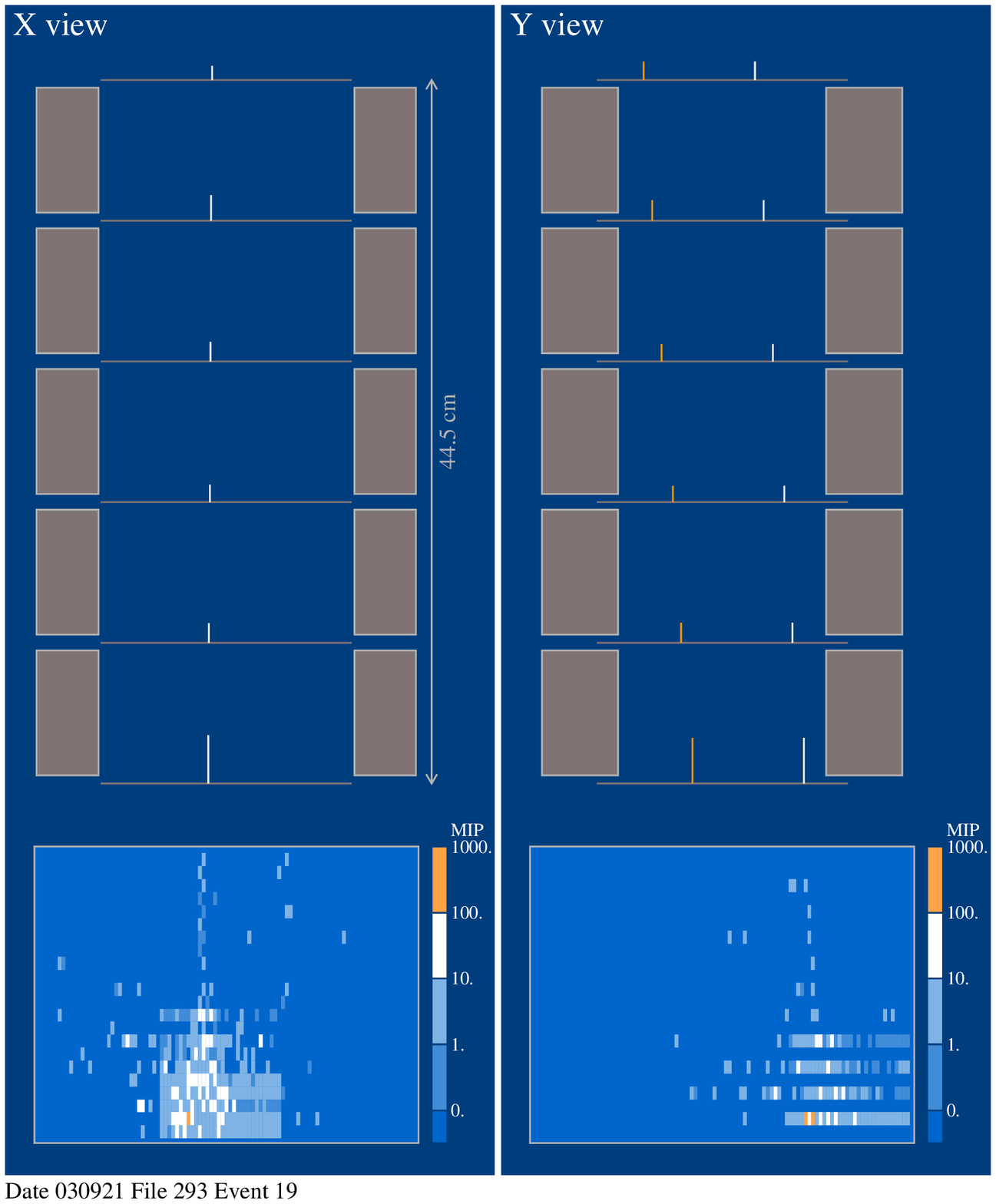, width=6.5cm}
\caption{An event display of a 50~GeV/c electron (left) and proton
(right) recorded at the CERN SpS facility. Hits in the tracking
system are shown (including ambiguities for the y-view) along with
activity in the calorimeter. The signals from the odd planes of
the y-view of the calorimeter were not read-out during this test.
One of the x-view planes was also not operational and was later
replaced.} \label{topo}
\end{center}
\end{figure}

The longitudinal and transverse segmentation of the calorimeter,
combined with the measurement of the particle energy loss in each
silicon strip, allows a high identification (or rejection) power
for electromagnetic showers. Electromagnetic and hadronic showers
differ in their spatial development and energy distribution in a
way that can be distinguished by the calorimeter. This is
demonstrated in figure~\ref{topo} which shows examples of an
electromagnetic shower induced by an electron (left) and an
interacting proton (right), recorded during tests with particle
beams at the CERN SpS facility. All incident particles have a
momentum of 50~GeV/c. The electron-hadron separation performance
of the calorimeter has been extensively studied~\cite{calosep}
and the calorimeter is found to have sufficient performance to reach the
primary scientific objectives of PAMELA, providing a proton rejection
factor of about 10$^5$ while keeping about 90\% efficiency in selecting
electrons and positrons. From simulations, an electron rejection factor
of about 10$^5$ in antiproton measurements (about 90\% antiproton
identification efficiency) is demonstrated.

\begin{figure}
\begin{center}
\epsfig{file=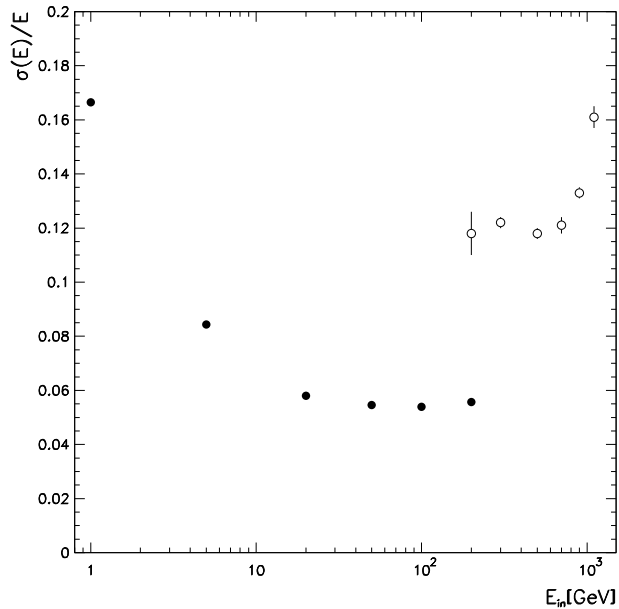, width=9cm} \caption{The energy dependence
of the energy resolution of the electromagnetic calorimeter. The
filled symbols are for normal operation (experimental data) and the open symbols are
for the self-trigger mode (simulations), described in section~\ref{sec:daq}.}
\label{eres}
\end{center}
\end{figure}

The calorimeter will also be used to reconstruct the energy of the
electromagnetic showers. This will provide a measurement of the
energy of the incident electrons independent from the magnetic
spectrometer, thus allowing a cross-calibration of the two
energy determinations. As shown in figure~\ref{eres}, the constant
term for the calorimeter energy resolution has been measured as
$\sim 5.5\%$ for electromagnetic showers generated by particles
entering the calorimeter within the acceptance of the tracking
system up to an energy of several hundred GeV.

The calorimeter is also equipped with a self-trigger capability, as discussed in
section~\ref{sec:trigger}.

\subsection{Shower tail catcher scintillator}
\label{s4}

\begin{figure}
\begin{center}
\epsfig{file=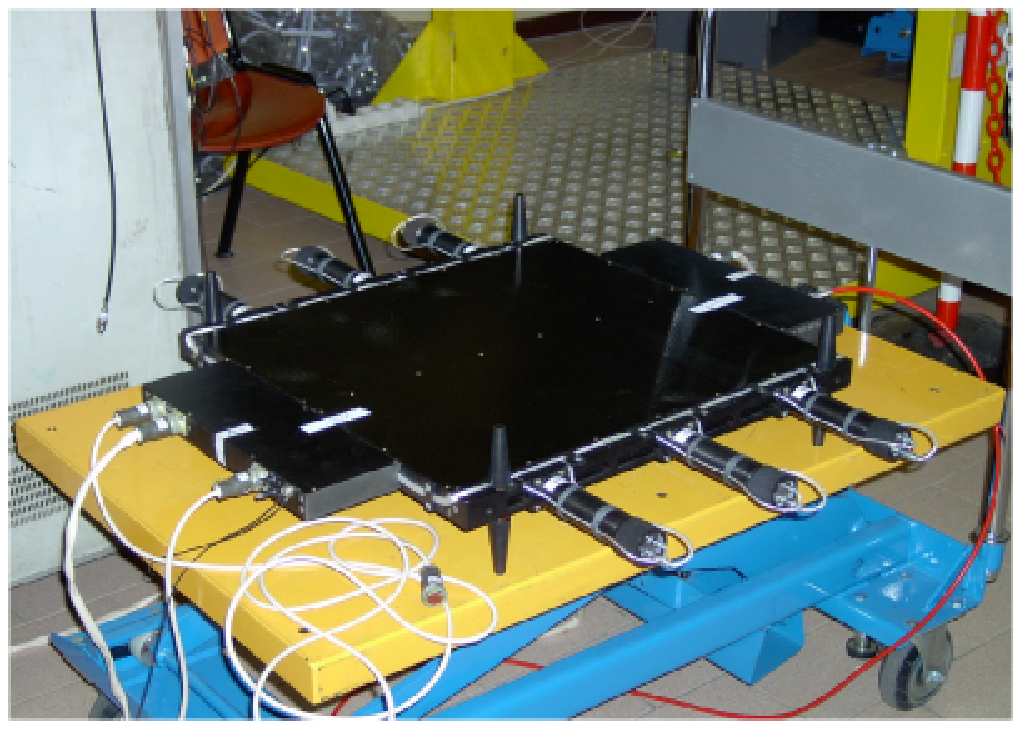, width=6.5cm}\hspace{2mm}\epsfig{file=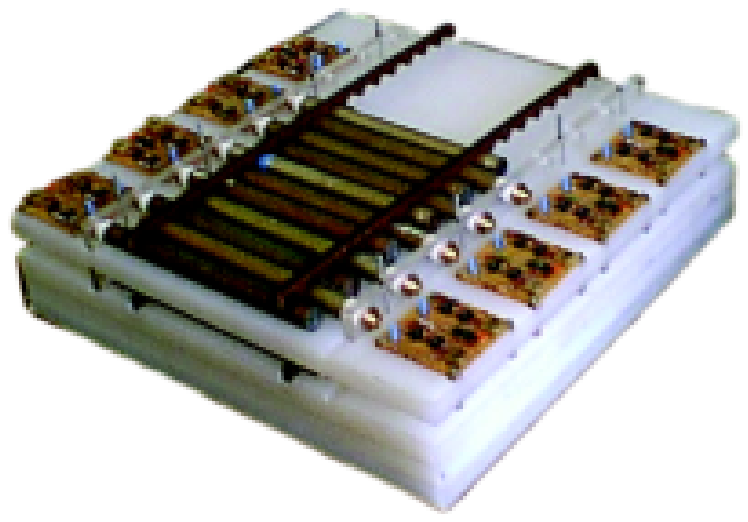, width=6.5cm}
\caption{Left: the shower tail catcher scintillator, S4, showing
the 6~PMTs used for read-out. The scintillator has dimensions (48$\times$48)~cm$^2$.
Right: The neutron detector
partially equipped with $^3$He proportional counters. The neutron detector covers an area of
($60 \times 55$)~cm$^2$.}
\label{nd_s4}
\end{center}
\end{figure}

The shower tail catcher scintillator (S4) improves the PAMELA
electron-hadron separation performance by measuring shower leakage
from the calorimeter. It also provides a high-energy trigger for
the neutron detector
 (described in the next section). This scintillator is placed directly beneath the calorimeter.
 It consists of a single square piece of 1~cm thick scintillator of dimensions (48$\times$48)~cm$^2$
 which is read out by six PMTs, as shown in figure~\ref{nd_s4}.

\subsection{Neutron detector}


The neutron detector complements the electron-proton
discrimination capabilities of the calorimeter. The evaporated
neutron yield in a hadronic shower is 10--20 times larger than
expected from an electromagnetic shower. The neutron detector is
sensitive to evaporated neutrons which are thermalised in the
calorimeter.
Joint analysis of the calorimeter and neutron
detector information are expected to allow allow primary electron energies to be
determined up to several~TeV.

The neutron detector~\cite{sto05} is located below the S4 scintillator and consists of 36 proportional counters,
 filled with $^3$He and surrounded by a polyethylene moderator enveloped in a thin cadmium layer to
 prevent thermal neutrons entering the detector from the sides and from below.
 The counters are stacked in two planes of 18 counters, oriented along the y-axis of the instrument.
 The size of the neutron detector is ($60 \times 55 \times 15$)~cm$^3$ and is shown in figure~\ref{nd_s4}.

\section{PAMELA data acquisition and trigger system}
\label{sec:daq}

\subsection{Data acquisition system}
\label{sec:daqdaq}

\begin{figure}
\begin{center}
\epsfig{file=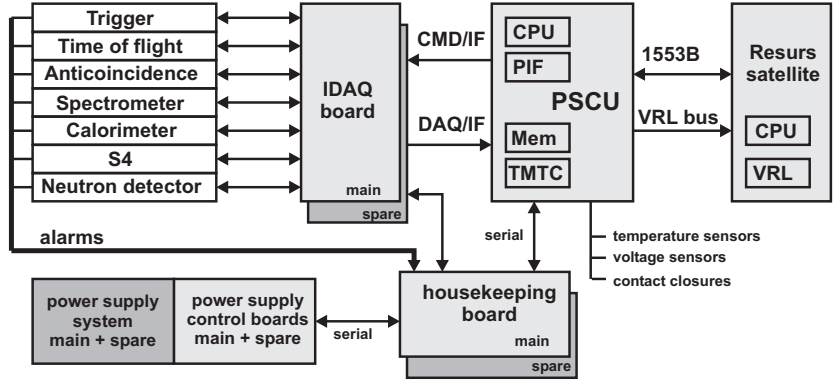, width=11cm} \caption{Scheme of
the PAMELA data acquisition system. The interfaces (IF) between
the IDAQ and the PSCU handle the data acquisition and transfer the
commands (CMD) to the IDAQ. The communication between PAMELA and
the spacecraft is handled by the 1553B bus and by the link to the
VRL module for data download to the Resurs memory. Adapted
from~\cite{pscucospar}.} \label{daq}
\end{center}
\end{figure}

A schematic overview of the PAMELA data acquisition (DAQ) system
is shown in figure~\ref{daq}. The PSCU (PAMELA Storage and Control
Unit) handles all slow controls, communication with the satellite,
data acquisition, storage and downlink tasks. The PSCU contains 4
subsystems: \begin{itemize}

\item[(i)] A processor module built around a CPU based on a ERC-32
architecture (SPARC v7 implementation) running the RTEMS real time
operating system at 24~MHz. The CPU is custom built by Laben and
is fully space qualified. There is no redundant back-up. Communication with the Resurs satellite
is realised via a standard 1553B data-bus;
\item[(ii)] Two
redundant 2~GByte mass memory modules. The modules include
latch-up detection, allowing operation to be transparently
switched to the safe module when a latch-up is detected;
\item[(iii)] A PIF (PAMELA interface board) that performs three
main tasks: communication with the IDAQ (Intermediate DAQ) system
through a DMA (dynamic memory access) controller, handling the
interface with the mass memory, and providing the interface with
the VRL (Very high-speed Radio Link) module of the satellite;
\item[(iv)] A TMTC (Telemetry and Control) board that handles the housekeeping operations
of PAMELA, such as alarm, temperature and voltage monitoring (once per second). Such monitoring is performed both directly (ADC
inputs and contact closure telemetries) and through a dedicated
housekeeping board that communicates through serial data links
with the subdetector read-out boards, with the IDAQ board and with
the power supply control boards.
\end{itemize}

Data acquisition from the subdetectors is managed by the IDAQ system at a rate of 2~MByte/s. Upon receipt of
a trigger, the PSCU initiates the IDAQ procedure to read out data
from the subdetectors in sequence. The resulting data are stored
in the PSCU mass memory. Several times a day, the data are
transferred to the satellite on-board memory via the 12~MByte/s
VRL bus where it is stored prior to downlinking to earth.
Approximately 15~GBytes are transferred to ground per day during 2-3 downlink sessions.

The PSCU automatically handles the flow of PAMELA physics tasks
and continuously checks for proper operation of the apparatus. At
boot, the PSCU manages the operation of the power supply system to
power up all subsystems, initializes all detectors and starts the data
acquisition cycle. In parallel, once per second the PSCU checks
the TMTC information on voltages and alarms. In case of abnormal
conditions the PSCU can perform a hardware reset of the whole
system or, if insufficient to solve the problem (e.g. in case of
electronics latch-up), powers down and then up PAMELA. The PSCU
also checks the temperature environment by reading dedicated
temperature sensors distributed in various locations around the
instrument. If the readings exceed predefined values (set with
dedicated commands from ground) the PSCU powers down PAMELA until
acceptable working conditions are reached. The PSCU also handles communication with the Resurs satellite CPU and
VRL system. Data is downloaded to the VRL upon receipt of a dedicated
command from the Resurs CPU. The scheduling of data downloads from the PAMELA mass memory
to the VRL hard disk system is defined from ground on a daily basis.

The PSCU organizes the data acquisition cycle in ``runs". A run is
defined as a continuous period of data taking in which the trigger
and detector configurations are constant. These configurations are
defined by the PSCU according to information stored in on-board
memory or received from ground. The duration of a run is
determined by the PSCU according to the orbital position (e.g.
inside radiation belts or South Atlantic Anomaly SAA or outside
these areas). The orbital position also dictates the trigger
configuration, as described in the following section. The orbital position is derived from the ``ascending node" notification issued by the Resurs CPU when the satellite
crosses the equator from the southern hemisphere to the northern hemisphere. From this position
information, the CPU extrapolates the entry time into high
radiation environments. This can be performed in three ways,
chosen from ground:
\begin{itemize}
    \item when the counting rate of the S1 scintillator exceeds a
given threshold (changeable from ground with dedicated command);
    \item according to fixed time periods conservatively chosen
and modifiable from ground;
    \item according to a table with crossing times in absolute
Moscow time\footnote{On a regular basis the Resurs CPU sends a time
synchronization command with the Moscow time to the PSCU. The
precision of this information is $\sim$ 1~s.} provided on a
bi-weekly basis from ground with a dedicated command.
\end{itemize}
Additionally, the PSCU can interrupt and close a run if anomalous
conditions that require action upon the subsystems (e.g. hardware
resets, etc.) are detected.

Periodically the PSCU calibrates the detectors, namely the anticounter system,
the tracker, the calorimeter and the S4 scintillator.
By default, the calibration is performed at the point of lowest cosmic-ray
trigger rate, i.e. the equator, upon receiving an ``ascending node"
notification from the Resurs CPU. The frequency of calibrations can be
modified from ground.

\subsection{Trigger system}
\label{sec:trigger}

The PAMELA trigger condition is defined by coincident energy
deposits in the scintillator ToF layers. Various configurations can be selected.
The default ones (the subscripts 1 and 2 refer to the upper and lower
layers in each ToF plane) used outside and inside radiation
environments are:
\begin{itemize}
\item \mbox{\textit{(S1$_{1}$ or S1$_{2}$) and (S2$_{1}$ or
S2$_{2}$) and (S3$_{1}$ or S3$_{2}$)}} outside radiation belts and
SAA; \item \mbox{\textit{(S2$_{1}$ or S2$_{2}$) and (S3$_{1}$ or
S3$_{2}$)}} inside radiation belts and SAA;
\end{itemize}
since, according to simulation, the radiation environment will
saturate the S1 counting rate but will not affect significantly
the S2 and S3 scintillators since they are more shielded.

These trigger configurations can be changed from
ground with dedicated commands to the PSCU. A total of 29~configurations have been implemented on the trigger board.  Various combination of {\it and} or {\it or} of the
scintillators layers with or without the calorimeter self-trigger
and S4 trigger (described below) are implemented. The PMTs can be masked on the trigger
board by the PSCU.

The calorimeter is equipped with a self-trigger capability. A
trigger signal is generated when a specific energy distribution is
detected in predetermined planes within the lower half of the
calorimeter. The sets of planes used in this configuration
can be changed with a dedicated command from ground. This allows
PAMELA to measure very high-energy ($\sim$300~GeV to $>$1~TeV)
electrons in the cosmic radiation. At present, very few
measurements have covered this energy range~\cite{kob99}. Since
these events are rare, it is important to have a large geometrical
factor. By requiring that triggering particles enter through one
of the first four planes and cross at least 10~radiation lengths,
the geometrical factor is $\sim$600~cm$^{2}$sr, i.e. about a
factor of 30 larger than the default PAMELA acceptance defined by the magnetic spectrometer. The
behaviour of the calorimeter in self-trigger mode has been studied
by means of simulations~\cite{boe02}. The simulated energy
resolution of the calorimeter in self-trigger mode is
approximately constant ($\sim$12\%) up to about 800~GeV, as shown in figure~\ref{eres}. At higher
energies the resolution decreases because of increasing
longitudinal leakage and saturation of the signal from the strips
(about 1000~mip). The choice of energy loss and activated planes
implemented in the calorimeter electronics to generate a trigger
signal has been taken to have the highest proton rejection while
keeping a trigger efficiency of better than 90\% for electrons of
energies higher than 300~GeV~\cite{boe02}. Combined with the
neutron detector information, the apparatus will be able to
cleanly identify very high-energy electrons. The neutron detector can also be triggered when an energy deposit exceeding 10~mip is detected in the S4 scintillator.

\begin{figure}
\begin{center}
\epsfig{file=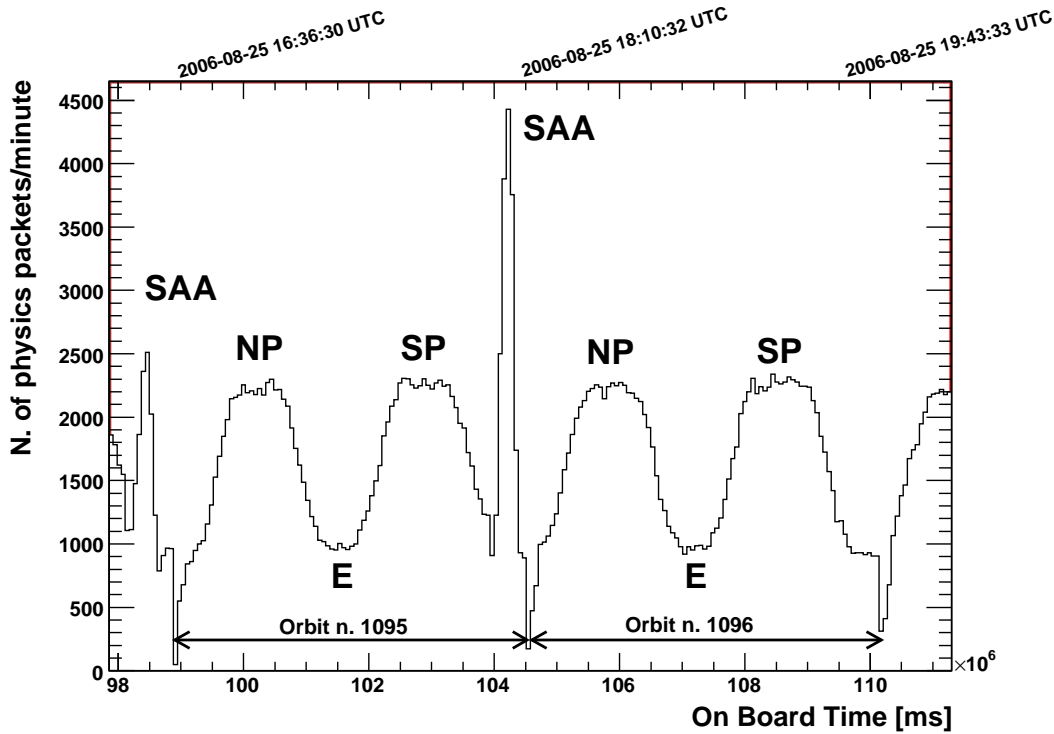, width=14cm} \caption{The
PAMELA trigger rate shown in events per minute evaluated during two consecutive orbits (period
$\sim$~94~minutes).  The trigger rate is strongly dependent on the
orbital position~: NP, North Pole; SP, South Pole; E, Equator; SAA , South Atlantic Anomaly (SAA).} \label{trig_rate}
\end{center}
\end{figure}

The trigger rate observed during typical orbits is shown in figure~\ref{trig_rate}. The maxima at $\sim$2200~events per minute ($\sim$35~Hz) correspond to passages over the polar regions (North Pole, NP and South Pole, SP) while the minima ($\sim$15~Hz) correspond to equatorial regions (E). The contribution from the
South Atlantic Anomaly (SAA) is clearly visible ($\sim$70~Hz, maximum). Note that data is taken in the SAA using the second default trigger configuration. The missing acquisition time after the peaks of the SAA
corresponds to the detector calibrations upon crossing the equator (about 1~minute in duration).

Dead and live times are monitored by two clocks that count the time during which the data acquisition system is busy or is waiting for a trigger, respectively. The dead time varies significantly over an orbit, due to the significant changes in trigger rate shown in figure~\ref{trig_rate}. Furthermore, if the satellite crosses the SAA the dead time increases. The dead time also depends on the trigger configuration. For an orbit not crossing the SAA the fractional dead time is approximately 26\%, i.e. the fractional live time is about 74\%. As discussed previously, an automatic procedure changes the trigger configuration when entering radiation environments thus reducing the trigger rate and, consequently, the dead time. Complete details of in-flight performance will be detailed in a future publication.

Large solar particle events (SPE) can lead to a high rate of particles hitting the top scintillator
(S1 in figure~\ref{pam}, top). Although very large events such as the one observed on 24$^{th}$ October 1989 could result in rates of $\sim$7~MHz on S11, most SPE will occur at solar minimum ($\sim$10 SPE are expected during a 3~year PAMELA mission) and will be of much smaller intensity. For example, a coronal mass ejection such as that of 24$^{th}$ September 1997 would result in a S1 rate of less than 100 Hz, much less than that encountered in the SAA ($\sim$ 1~kHz). For large events the automatic trigger selection procedure would switch to a configuration without the S1 detector, as currently happens during passages through the SAA. After this, a specific trigger configuration suited to the size and expected temporal evolution of the event can be selected from ground.

If the amount of event data exceeds the storage dedicated to
PAMELA on-board the Resurs satellite or the daily
downlink limit, an on-line event selection is provided
by a second level trigger. The second level trigger is not
normally activated and must be activated via an uplinked command
from ground. Information from the CAS anticoincidence system is
used to reject ``false" triggers (see section 2.3) and information
from the calorimeter is used to reduce the impact of particles
backscattered from the calorimeter. The second level trigger is
described in detail elsewhere~\cite{level2}.

\section{The Resurs DK1 satellite}
\label{sec:dk1}

The Resurs DK1 satellite is manufactured by the Russian space
company TsSKB Progress to perform multi-spectral remote sensing of
the earth's surface and acquire  high-quality images in near
real-time. Data delivery to ground is realised via a high-speed
radio link.

\begin{figure}
\begin{center}
\epsfig{file=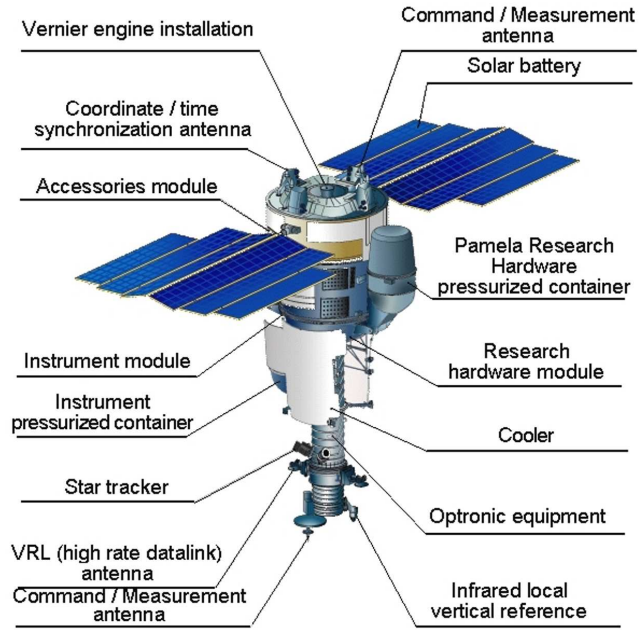, width=9cm} \caption{A sketch of the
Resurs DK1 satellite which hosts the PAMELA experiment in a
Pressurized Container (shown in the data-taking position). The satellite has a height of 7.4~m.}
\label{resurs-pc}
\end{center}
\end{figure}

The satellite is presented in figure~\ref{resurs-pc}, has a
mass of $\sim$6.7~Tonnes and a height of 7.4~m. The solar array
span is $\sim$14~m.
The satellite is three-axis stabilized with an axis orientation accuracy of 0.2~arcmin and an angular velocity stabilization accuracy of 0.005$^\circ$/s.
The orbital altitude varies between
350~km and 600~km at an inclination of 70$^\circ$. The design
lifetime is three years.

\begin{figure}
\begin{center}
\epsfig{file=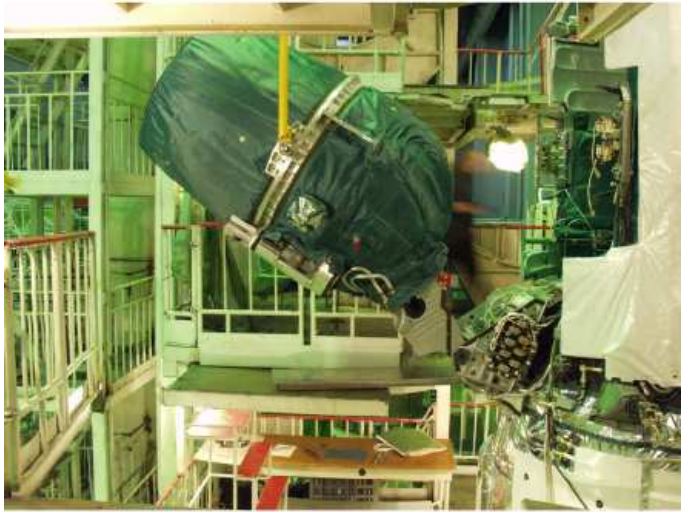, width=9cm}
\caption{Tests of the PAMELA Pressurized Container during orbital operations (May 2002). The body of the Resurs DK1 satellite can be seen to the right of the picture.}
\label{pc_test}
\end{center}
\end{figure}

PAMELA  is mounted in a dedicated Pressurized Container (PC)
attached to the Resurs DK1 satellite. During launch and orbital
manoeuvres, the PC is secured against the body of the satellite.
During data-taking it is swung up to give PAMELA a clear view into
space. The container is cylindrical in shape and has an inside
diameter of about 105~cm, a semi-spherical bottom and a conical
top. It is made of an aluminium alloy, with a thickness of 2~mm in
the acceptance of PAMELA. Figure~\ref{pc_test} shows tests of the
PC tilting mechanism performed in May 2002 at the TsSKB Progress
facility in Samara. The movement of the PC from the parked to the
data-taking position was tested in simulated weightless conditions.

\subsection{NTs OMZ ground segment}

The ground segment of the Resurs DK1 system is located at the
Research Center for Earth Operative Monitoring (NTs~OMZ) in
Moscow, Russia~\cite{ntsomz}. This forms part of the Russian Space Agency
(Roskosmos) ground segment designed for acquiring, recording,
processing and distributing data from remote sensing systems in
space.

The reception antenna at NTs~OMZ is a parabolic reflector of 7~m
diameter, equipped with an azimuth-elevation rotation mechanism,
and has two frequency multiplexed radio channels. The Resurs DK1
radio link towards NTs~OMZ is active 2-3 times a day. The average
volume of data transmitted during a single downlink is currently
$\sim$6 GBytes, giving a total of 15~GBytes/day. Data received from
PAMELA are collected by a data-set archive server. The server
calculates the downlink session quality (the error probability per
bit) and faulty downlink sessions can be assigned for
retransmission up to several days after the initial downlink. The downlinked data are transmitted
to a server dedicated to data processing for instrument monitoring
and control, and is also written to magnetic tape for long-term
storage. All such operations are automatized to minimize the time
delay between the data reception and the extraction of monitoring
information.

After this first level of data analysis, both raw and preliminary processed data are moved through a normal internet line to the main storage centre in Eastern Europe, which is located at MePHI
(Moscow, Russia). From here, GRID infrastructure is used to move raw and first level processed data to
the main storage and analysis centre of the PAMELA Collaboration,
located at CNAF (Bologna, Italy), a specialized computing centre
of INFN. Here data are accessible to all various institutions
within the PAMELA collaboration.

\section{Qualification tests}
\label{sec:qual}

Space-borne apparatus must maintain a high level of performance and
stability throughout the mission duration in the harsh environment
of space. The mechanical design must be such that the payload and
satellite withstand the significant shocks and vibrations of the
launch. The extremes of temperature that may be encountered in
space require that the thermal and mechanical designs be such
that the sensitive components maintain excellent stability over a broad range of temperatures.
The radiation environment in space is a major consideration in the design of
electronic circuitry. All chosen components must be tested for radiation tolerance prior to use.
Electromagnetic interference (EMI) from electronic devices must be
minimized by the use of different types of filters and shielded cables.

In this section the steps taken to qualify PAMELA for operation in space are reviewed.

\subsection{Radiation tolerance}

In orbit all on-board electronic devices will be subject to
the passage of ionizing particles, which can degrade their
performance and eventually lead to their permanent damage or loss
of functionality. Since malfunctioning components cannot be replaced
once the instrument is in orbit, all critical devices must either
be already space qualified, or tested for radiation tolerance
before use.


For economic, performance and power consumption reasons, most of
the PAMELA electronic components are ``off-the-shelf" commercial
products. Radiation tolerance tests therefore had to be carried
out before their integration into electronic boards. A selection
of electronic components have been tested under gamma and heavy
ion beams during the construction phase of the PAMELA subsystems.


As an example, the DSP and FPGA chips used through-out
the PAMELA data acquisition system were extensively tested in the
period 2000-2002, using heavy-ion beams. The tests were performed
at GSI in Darmstadt (Germany), and JINR in Dubna (Russia). At GSI
the devices were exposed to beams of $^{131}$Xe and $^{238}$U, in
the energy range 100-800~MeV/n. Different incidence angles allowed
different doses to be achieved. At JINR slow beams of $^{24}$Mg at
150~MeV/n were used, in order to maximize the energy transfer to
the components under test. Test results have been published
elsewhere~\cite{bos03}.

It is not expected that Solar Particle Events pose a hazard to PAMELA. However, in case of very large events PAMELA will be switched off in the time occurring between observation of the flare at the Sun and the particles reaching earth.


\subsection{Mechanical Qualification}

The mechanical and thermal space qualification tests of the PAMELA
instrument were performed in the years 2002-2003. In order to
perform such tests, a mock-up of the entire instrument,
Mass-Dimensional and Thermal Model (MDTM), was manufactured. The
MDTM reproduces the geometrical characteristics of PAMELA (e.g.
dimensions, total mass, center of gravity, inertial moments) and
the basic thermal behaviour. All particle detectors in the MDTM
were simulated by dummy aluminium boxes. The electronics systems
were non-functional and only reproduced the power consumption of
each subsystem.

\begin{figure}
\begin{center}
\epsfig{file=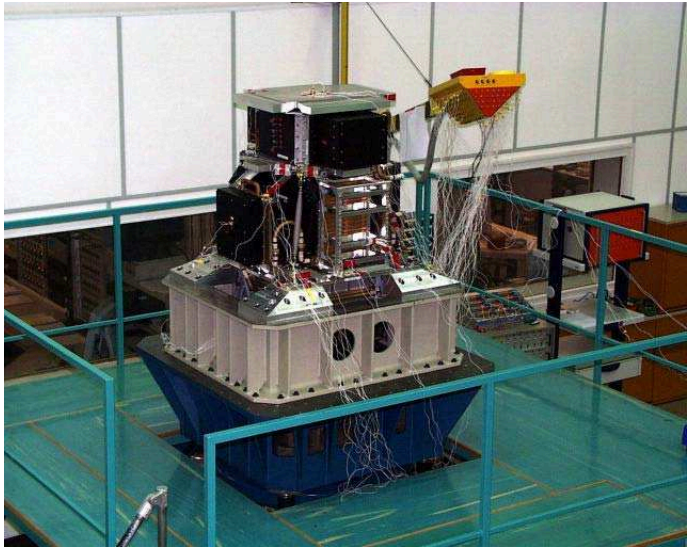, width=9cm}
\caption{The PAMELA MDTM on the shaker system in IABG (August 2002).}
\label{iabg}
\end{center}
\end{figure}

In order to ensure that no damage occur to PAMELA or the
spacecraft during any of the different operational phases of the
mission (transport, launch, orbital operations, unlocking of the
Pressurized Container, flight), the MDTM was exposed to
vibration spectra at mechanical loads exceeding those expected during the mission.
The MDTM vibration tests were performed at IABG Laboratories (Munich, Germany) in August
2002, as shown in figure~\ref{iabg}.
During the test it was verified that structural integrity was
maintained and that there was no change in the dynamic behaviour of
MDTM (using resonance searches). The MDTM structure was subjected
to the required vibration loads along three orthogonal axes.
Additional transport, vibration and shock tests of the MDTM whilst
integrated into the Pressurized Container were performed at the
TsSKB-Progress Testing Center in May~2003. Additional
information about PAMELA mechanical space qualification can be
found in \cite{sparvoli1}.

\subsection{Thermal Qualification}

The PAMELA thermal cooling system consists of a 8.6~m long pipe that joins 4~radiators
and 8~flanges connected throughout the PAMELA detector system. The task of this system is to dissipate the heat produced
by the PAMELA subsystems and transfer it into the spacecraft, where a custom designed thermal control system is located. This transfer
is performed by means of a heat-transfer fluid pumped by Resurs satellite
through the PAMELA pipelines. The total heat release of PAMELA cannot
exceed 360~W.

Thermal and vacuum tests of the PAMELA MDTM were performed in the
laboratories of TsSKB-Progress in April~2003. Six thermal modes of
operation were implemented, where the three relevant parameters
which regulate the instrument thermal behaviour (PAMELA power
consumption, external heat flows and heat-transfer fluid
temperature and flow rate) were varied between the design extrema
to simulate in-flight operations. Each mode persisted until a
steady state condition was reached. As an example, a test
simulating an interruption in the flow of the heat-transfer fluid
due to a malfunction was interrupted after 3~hours when the PAMELA MDTM
reached a temperature of $\sim$60$^\circ$C.

\begin{figure}
\begin{center}
\epsfig{file=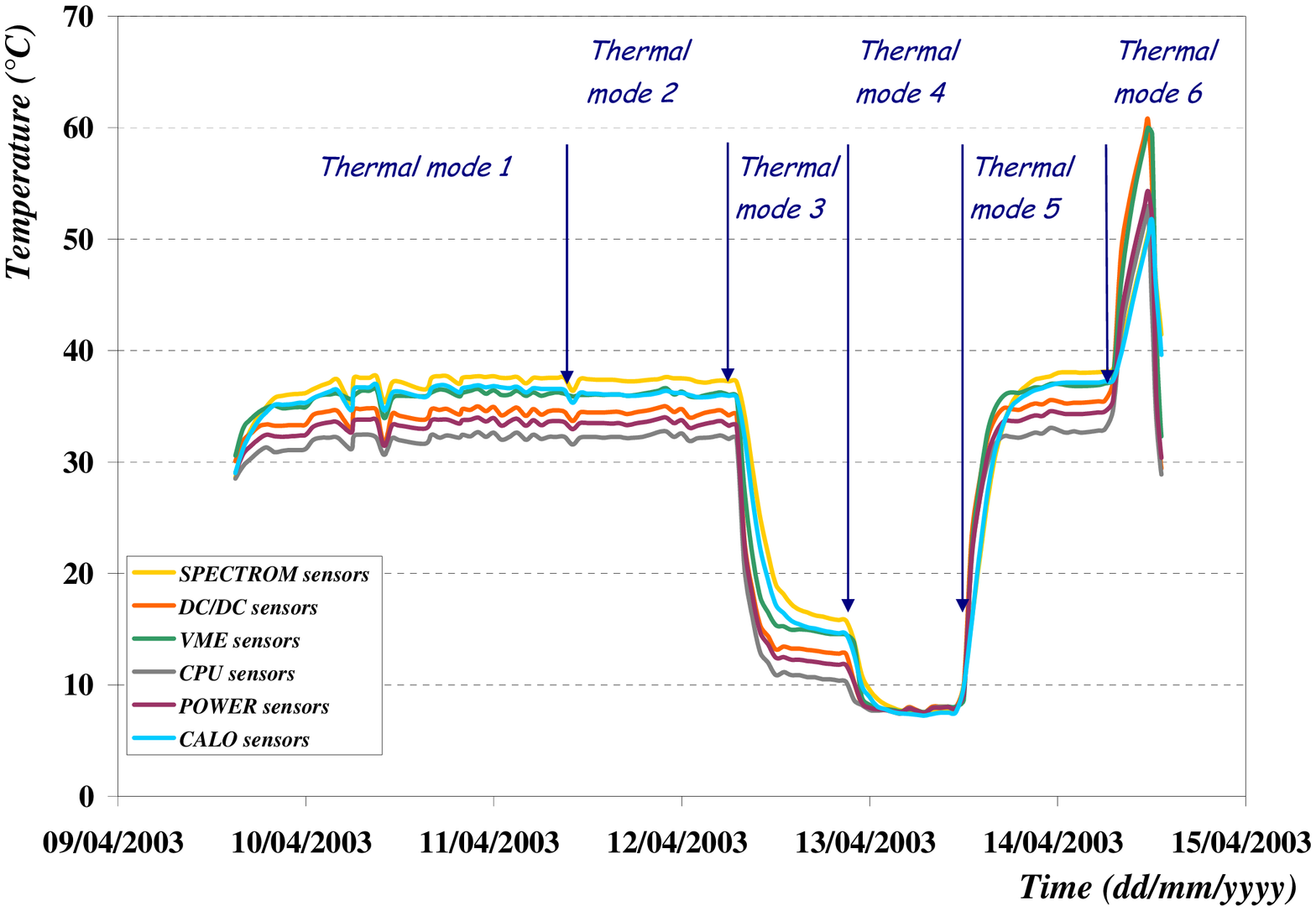, width=12cm} \caption{Results of the
PAMELA thermal qualification tests. Temperatures in different
subsystems are shown during the execution of the 6 different
thermal modes. The temperature remained always between acceptable
limits (5$^\circ$C - 40$^\circ$C) except for thermal mode number~6 where a stop in the heat
transfer fluid was simulated.} \label{dallas}
\end{center}
\end{figure}

The qualification test of the PAMELA thermal system showed that
all parameters of the system stayed within the design limits
(5$^\circ$C - 40$^\circ$C). The test shows that during the Resurs DK1 orbit the expected operating temperature
range of PAMELA will vary between 7$^\circ$C for the coldest systems and 38$^\circ$C
for the warmest ones, as shown in figure~\ref{dallas}.

Additional information about PAMELA thermal space qualification
can be found in \cite{sparvoli1}.

\subsection{Electrical tests}

\begin{figure}
\begin{center}
\epsfig{file=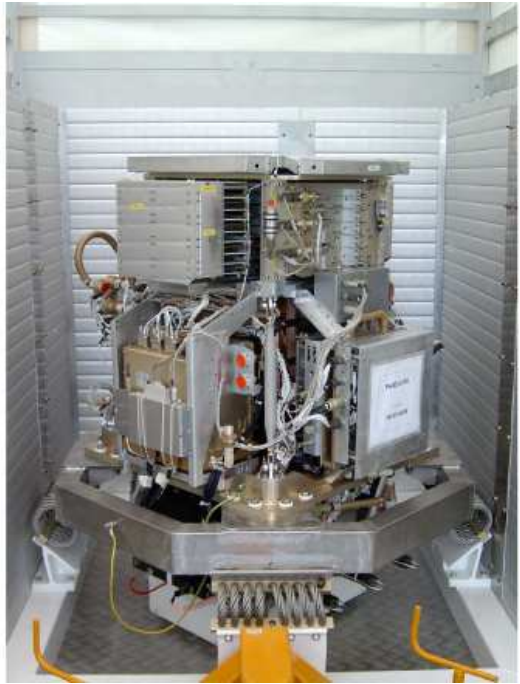, width=7cm} \caption{The PAMELA
Technological Model during transportation from Rome to the
TsSKB-Progress plant (April 2004).} \label{techno}
\end{center}
\end{figure}

To perform tests of the electrical interface between PAMELA and the
spacecraft, a second mock-up of the PAMELA instrument was
assembled. This ``Technological Model" was an exact copy of
the Flight Model from the point of view of electrical connections to the satellite and for the readout electronics boards, with the particle detectors substituted by dummies.
The Technological Model was shipped to TsSKB-Progress in April~2004 (see figure~\ref{techno}).
The task of the Technological Model was to thoroughly test the
electrical interface to the Resurs DK1 satellite. In addition, it
was used to check that the residual magnetic field from the PAMELA
spectrometer did not interfere with the Resurs instrumentation.
These complex tests proceeded in phases. A first test was
performed in Rome in December~2003, with the satellite emulated by
a Ground Support Equipment (EGSE) system. A second test started in
May~2004 at TsSKB-Progress and verified the powering procedures.
In October~2004 the PAMELA Technological Model was fully
integrated into the Resurs DK1 to complete all remaining tests.

\section{Physics Performance}
\label{sec:ground}

\subsection{Beam tests}

Between July 2000 and September 2003, the PAMELA subsystems were
periodically exposed to particle beams at the CERN PS and SPS
facilities. Electron and proton beams were used with energies in
the 10's - 100's GeV range. Results from these tests are described
in section~\ref{sec:pamela}.


\subsection{Ground data}

\begin{figure}
\begin{center}
\epsfig{file=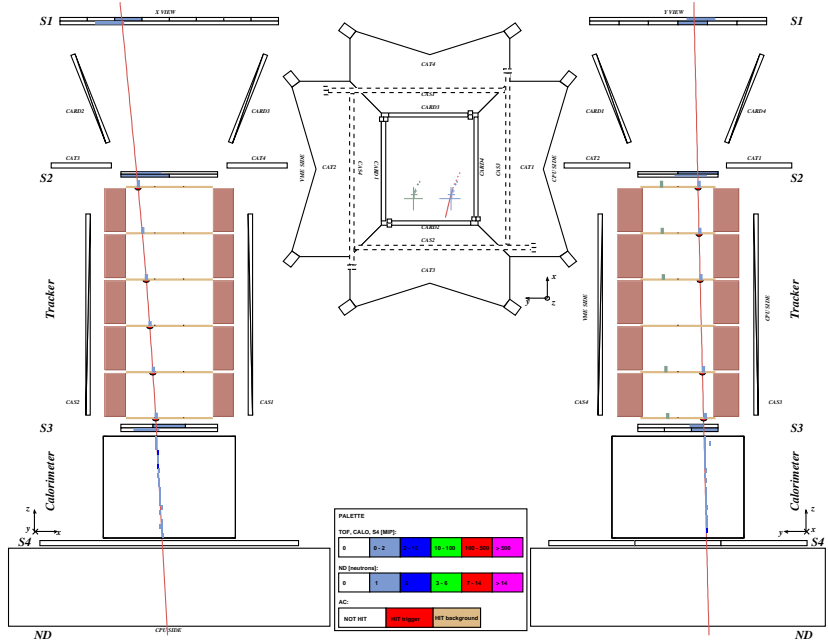, width=11cm} \caption{The event display of a
1.5~GeV/$c$ $\mu^{-}$ from ground data. On the left (right) the x,
bending view (y, non-bending view) of PAMELA are indicated. A plan view of
PAMELA is shown in the centre. The signals as detected by PAMELA
detectors are shown along with the particle direction (solid
lines) reconstructed by the fitting procedure of the tracking
system.} \label{muon}
\end{center}
\end{figure}

\begin{figure}
\begin{center}
\epsfig{file=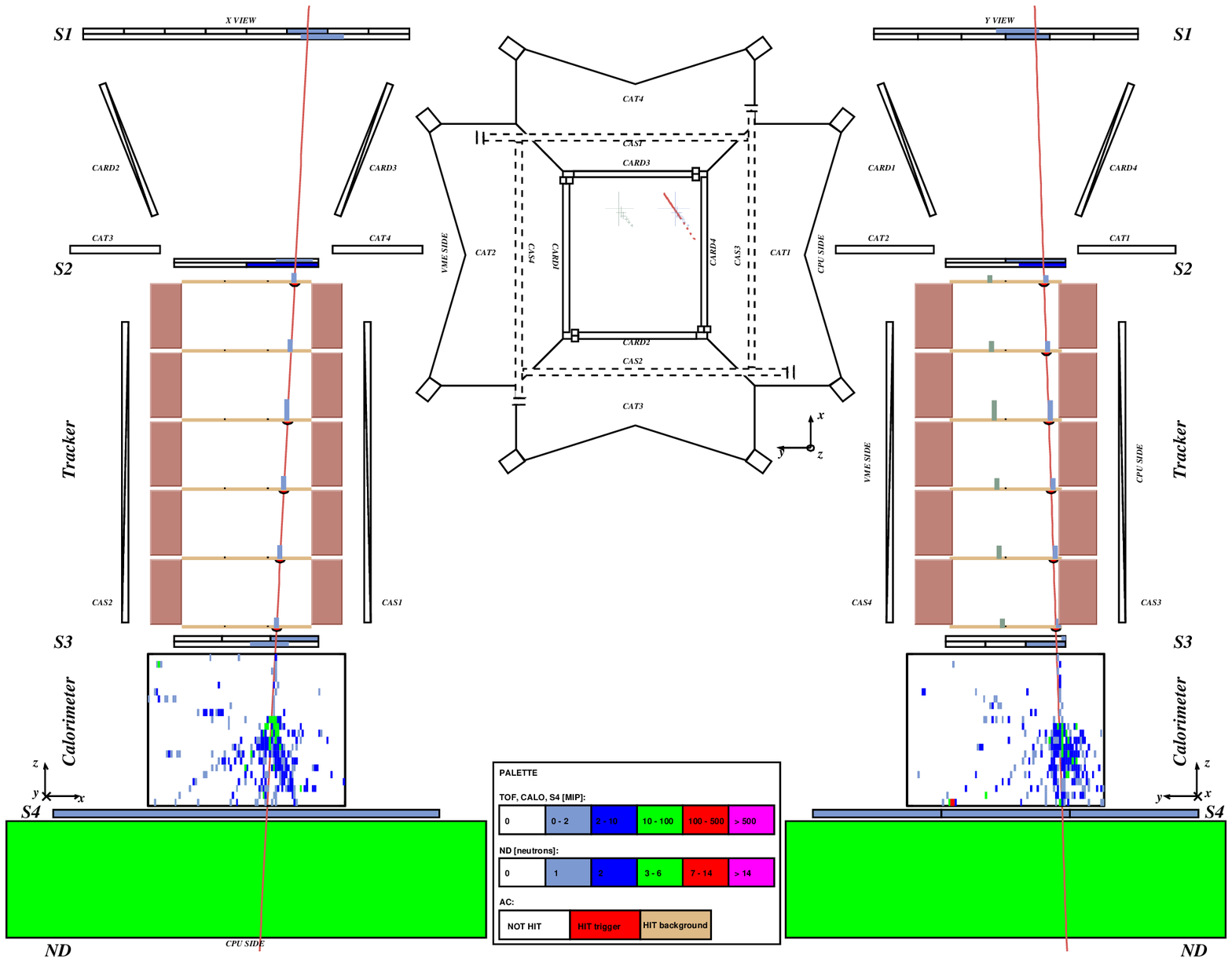, width=11cm} \caption{The event display of
a 67~GeV/$c$ hadron from ground data. On the left (right) the x,
bending (y, non-bending view) of PAMELA are indicated. A plan view of PAMELA is
shown in the centre. The signals as detected by PAMELA detectors
are shown along with the particle direction (solid lines)
reconstructed by the fitting procedure of the tracking system.}
\label{prot}
\end{center}
\end{figure}

Prior to delivery to Russia, the PAMELA instrument was assembled
at the INFN laboratories of Roma Tor Vergata, Rome, Italy. The system was tested with cosmic rays over a period of
several months. Figures~\ref{muon} and \ref{prot} show two cosmic
ray events recorded in Rome.
The first is a 1.5~GeV/$c$ negatively charged particle,
with high probability of being a $\mu^{-}$ considering the clean
non-interacting pattern in the calorimeter. The second is a
67~GeV/$c$ particle with a hadronic interaction in the
calorimeter, consistent with a proton. All PAMELA detectors are
shown in the figures along with the signals produced by the
particles in the detectors and derived information. Highly detailed information is provided for each
cosmic-ray event. The solid lines indicate the tracks
reconstructed by the fitting procedure \cite{van05} of the magnetic spectrometer.
The figures show also the ``ghost" hits due to the common readout of the 2~silicon sensors of the same ladder in the non-bending projection. This ambiguity is solved with the help of track fitting procedure and with a consistency check with the other PAMELA subdetectors.

\subsection{In-orbit performance}

\begin{figure}
\begin{center}
\epsfig{file=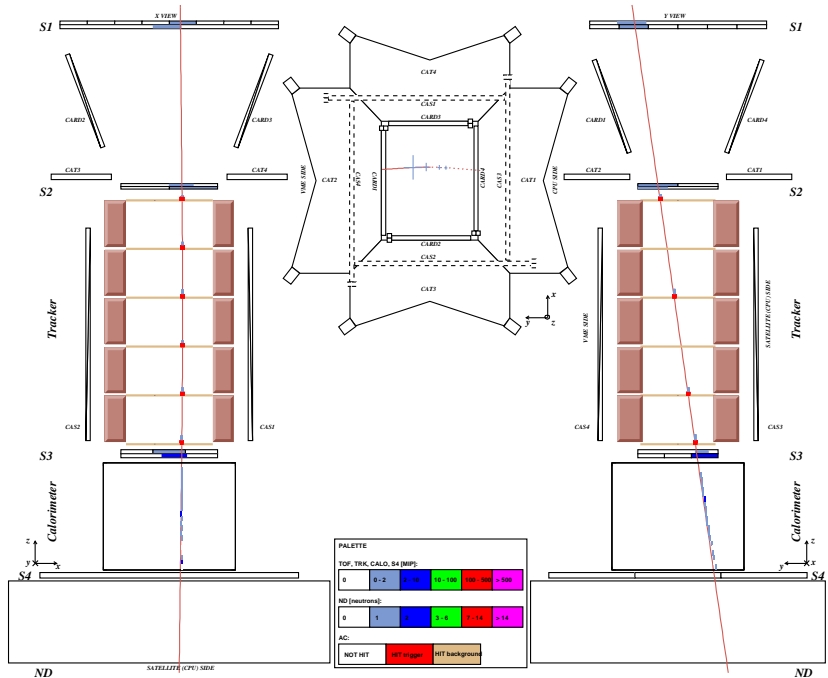, width=11cm} \caption{The event
display of a $\sim$3~GV non-interacting proton from flight data. On
the left (right) the x, bending view (y, non-bending view) of PAMELA are
indicated. A plan view of PAMELA is shown in the centre.}
\label{flight1}
\end{center}
\end{figure}

\begin{figure}
\begin{center}
\epsfig{file=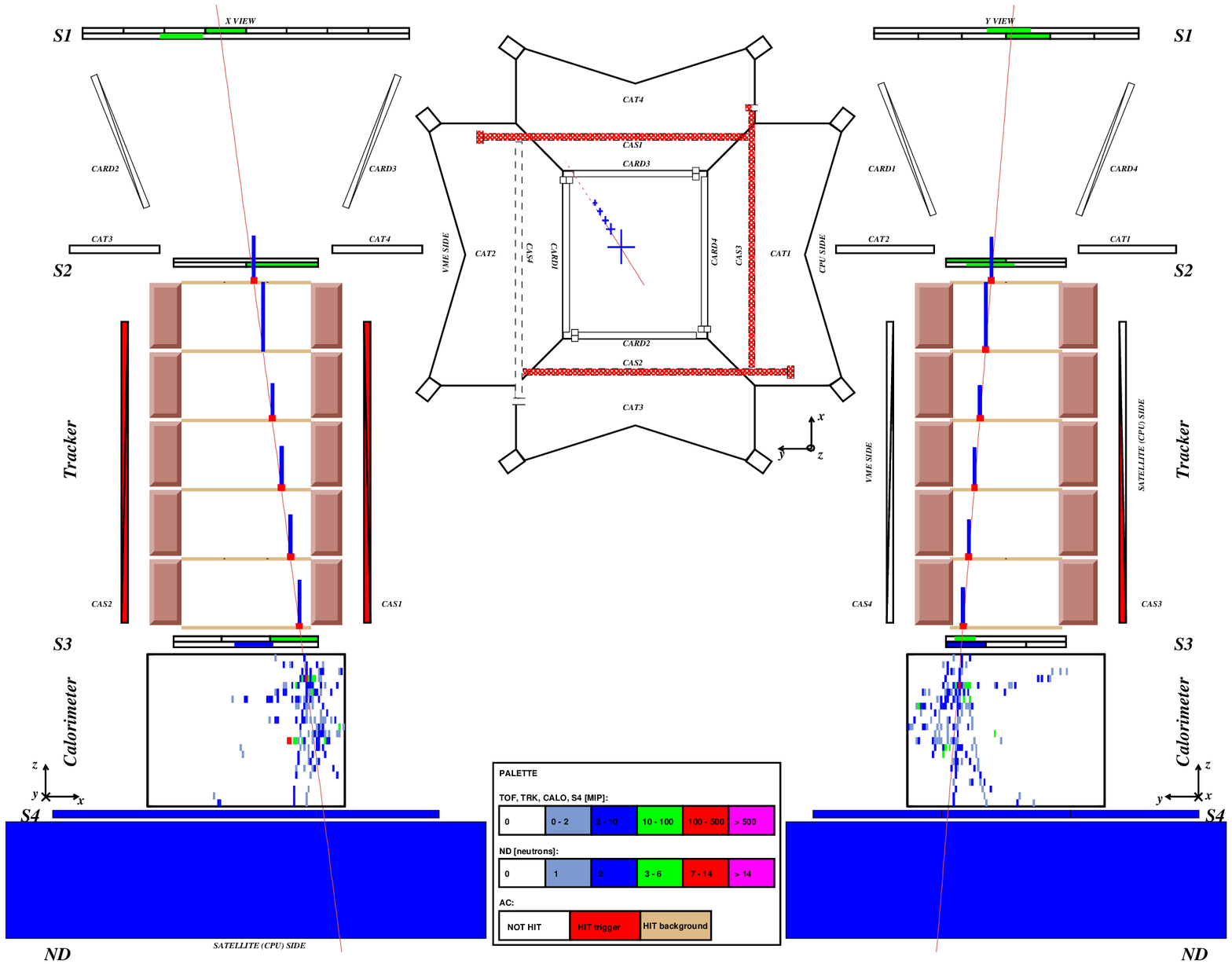, width=11cm} \caption{The event
display of a $\sim$13~GV interacting helium nucleus from flight
data. On the left (right) the x, bending (y, non-bending view) of PAMELA are
indicated. A plan view of PAMELA is shown in the centre. Note the increased energy deposit in the silicon tracker planes (denoted by the vertical bars) compared to figure~\ref{flight1}. The activity in the anticounter system is probably due to secondary particles backscattered from the calorimeter.}
\label{flight2}
\end{center}
\end{figure}

PAMELA was successfully launched on June 15$^{th}$ 2006 and was
first switched-on on the 20$^{th}$ of June. After a brief period of commissioning
PAMELA has been in a continuous data-taking mode since July 11$^{th}$.
Data downlinked to ground show that the entire instrument is working as expected. Figure~\ref{flight1} shows a 3~GV non-interacting proton recorded in-orbit while figure~\ref{flight2} shows a 13~GV helium nucleus interacting in the calorimeter.
The in-orbit performance of PAMELA  will be discussed in future publications.


\section{Conclusions}

PAMELA is a multi-purpose satellite-borne
apparatus designed to study charged particles in the cosmic
radiation with a particular focus on antiparticles (antiprotons
and positrons). The energy range over which observations are made and the foreseen statistics mean that PAMELA stands to deliver results of great scientific relevance in several fields of cosmic ray research. To ensure reliable operation in space extensive space qualification tests of the PAMELA detector systems, electronics and mechanical structures were performed prior to launch. The performance of the individual detector components and the PAMELA system as a whole were also verified at particle beam facilities.
The PAMELA instrument was launched into orbit from the Baikonur Cosmodrome on-board a Resurs DK1 satellite on June 15$^{th}$ 2006. All systems have been observed to operate as expected and scientific data analysis is now on-going.

\section{Acknowledgments}

The PAMELA mission is sponsored by the Italian National Institute of
Nuclear Physics (INFN), the Italian Space Agency (ASI), the
Russian Space Agency (Roskosmos), the Russian Academy of Science,
the German Space Agency (DLR), the Swedish National Space Board
(SNSB) and the Swedish Research Council (VR).
PAMELA is the result of a collaborative work that has lasted for several years. We gratefully acknowledge the contributions from the Italian companies Carlo Gavazzi Space, Laben, CAEN, Kayser, Forestal and Aerostudi. We also would like to thank the scientific laboratories and test facilities that assisted the PAMELA team during the qualification phases~: CERN (Geneva, Switzerland), JINR (DUBNA, Russia), GSI (Darmstadt, Germany), IABG (Munich, Germany), ENEA-CASACCIA (Rome, Italy) and GALILEO (Florence, Italy). We also thank the following engineers and technicians for their valuable contribution to the project~: L.~Andreanelli,
E.~Barbarito, A.~Bazarov, F.~Ceglie, S.~Ciano, C.~Fiorello, M.~Franco, A.~Gabbanini, E.~Gaspari, M.~Grandi, M.~Lundin, G.~Mazzenga, M.~Mongelli, R.~M\"{o}llerberg, O.~Panova, P.~Parascandolo, G.~Passeggio, G.~Pontoriere, E.~Reali, R.~Rocco, S.~Rydstr\"{o}m, A.~Sedov, B.~Talalaev, M.~Tesi, E. Vanzanella, and V.~Zotov. We thank the data reception centre, NTs OMZ, for its first class preparation of the PAMELA data receiving station in Moscow. Finally, we would like to express our gratitude to TsSKB-Progress, who carefully followed the PAMELA mission during all its phases and made the PAMELA launch a reality.


\end{document}